\documentclass[aps,twocolumn,prx,amsfonts,showpacs,superscriptaddress,longbibliography]{revtex4-2}
\usepackage{graphicx}
\usepackage{bm}
\usepackage{hyperref}
\usepackage{amsmath}
\usepackage{amssymb}
\usepackage[table]{xcolor}
\usepackage{array}
\usepackage{multirow}
\usepackage[caption=false]{subfig}
\usepackage{physics}
\usepackage{soul}
\usepackage{amsthm}
\usepackage{bbm}
\usepackage{multirow}
\usepackage[colorinlistoftodos]{todonotes}
\usepackage[scr=boondoxo,frak=boondox,scrscaled=1.05]{mathalfa}
\usepackage{amsmath}

\def\ket#1{|#1\rangle}
\def\braket#1#2{\langle#1|#2\rangle}

\def\r{{\boldsymbol{r}}}
\def\x{{\boldsymbol{x}}}
\def\k{{\boldsymbol{k}}}
\def\p{{\boldsymbol{p}}}
\def\v{{\boldsymbol{v}}}
\def\q{{\boldsymbol{q}}}
\def\A{{\boldsymbol{A}}}

\def\R{{\boldsymbol{R}}}
\def\G{{\boldsymbol{G}}}
\def\Q{{\boldsymbol{Q}}}
\def\X{{\boldsymbol{X}}}
\def\a{{\boldsymbol{a}}}
\def\b{{\boldsymbol{b}}}
\def\L{{\boldsymbol{L}}}
\def\J{{\boldsymbol{J}}}

\newcommand{\bk}{\bm {k}}

\newcommand{\br}{\bm {r}}
 
\usepackage{ulem}
\begin{document}

\title{Topological superconductivity with emergent vortex lattice in twisted semiconductors}

\author{Daniele Guerci}

\affiliation{Department of Physics, Massachusetts Institute of Technology, Cambridge, MA-02139, USA}

\author{Ahmed Abouelkomsan}

\affiliation{Department of Physics, Massachusetts Institute of Technology, Cambridge, MA-02139, USA}

\author{Liang Fu}
\affiliation{Department of Physics, Massachusetts Institute of Technology, Cambridge, MA-02139, USA}

\begin{abstract}

The coexistence of superconductivity and fractional quantum anomalous Hall (FQAH) effect has recently been observed in twisted MoTe$_2$ and theoretically demonstrated in a model of repulsively interacting electrons under an emergent magnetic field arising from the layer pseudospin texture in moir\'e superlattice~\cite{GuerciAbouelkomsan2025}. Here, we show that this superconducting state is a chiral $f$-wave superconductor hosting an array of {\it double} vortices, which are induced by the emergent magnetic field with $h/e$ flux quanta per moir\'e unit cell. This superconducting vortex lattice state is topological and features Chern number $-1/2$, giving rise to an half-integer thermal Hall conductance. Our theory provides a common mechanism and unified understanding of FQAH and topological superconductivity, with a rich phase diagram controlled by the spatial modulation of the emergent magnetic field. 
 
\end{abstract}
\maketitle
\date{\today}

\section{Introduction}

Magnetic fields induce a wealth of quantum phenomena in solids. Two paradigmatic examples are: (1) the fractional quantum Hall effect~\cite{Tsui1982,Laughlin1983}, which arises when two-dimensional electron gases are subjected to strong external fields; (2) the Abrikosov vortex lattice in superconductors under a magnetic field~\cite{abrikosov1957magnetic}, with each vortex carrying a superconducting flux quantum $h/2e$. Interestingly, the physical phenomena that are realized under externally applied magnetic fields can, in principle, be also realized in the presence of \textit{emergent} magnetic fields through various mechanisms. 
For example, quantum particles moving in a non-coplanar spin texture acquire a real-space Berry phase, equivalent to an emergent, magnetic field that is determined by the skyrmion density~\cite{Bruno2004}. The emergent magnetic field can give rise to flat Chern bands that mimic Landau levels~\cite{Paul_2023}.

\begin{figure}
    \centering
    \includegraphics[width=.8\linewidth]{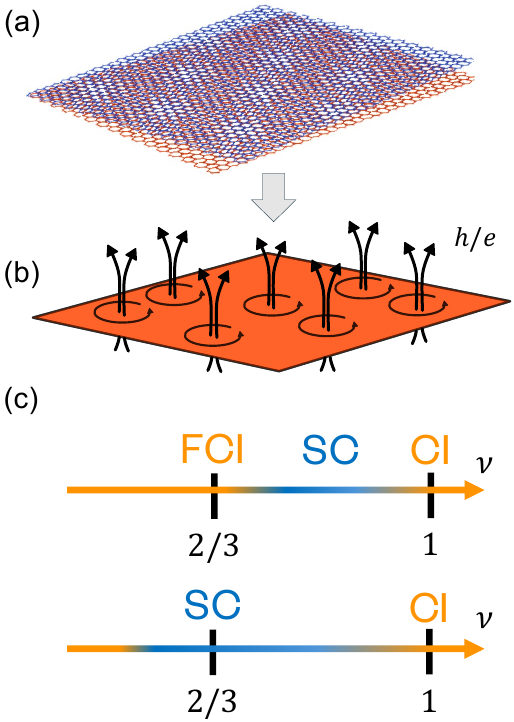}
    \caption{The twisted heterostructure (a) hosts a skyrmion lattice resulting in an inhomogeneous magnetic field (b) with two superconducting flux quanta per unit cell. For weak inhomogeneity of the emergent magnetic field (top image in panel (c)), anomalous Hall metal and  superconducting correlations develop for fillings $2/3<\nu<1$. Increasing the inhomogeneity drives a first-order transition from the fractional Chern insulator to a superconducting phase $\nu=2/3$, allowing superconductivity to extend to lower fillings. 
    }
    \label{fig:skyrmion+VL}
\end{figure}

This mechanism underlies the recently observed fractional quantum anomalous Hall (FQAH) effect in twisted bilayer transition metal dichalcogenide (TMD) MoTe$_2$ ($t$MoTe$_2$)~\cite{Devakul_2021,Li2021PRR, crepel2023anomalous,  Cai2023,zeng2023thermodynamic, xuParkObservationFractionallyQuantized2023,PhysRevX.13.031037}. 
Here, interlayer tunneling and the intra-layer moir\'e potentials act as a \textit{layer} pseudospin Zeeman field~\cite{FW_PRL_2019}, which forms a skrymion lattice and creates an emergent magnetic field with one flux quantum per unit cell~\cite{Paul_2023,MoralesDuran2024,Shi2024,Kolar2024,Li2025}. 
While emergent magnetic fields are equal and opposite for $K$-valley spin-$\uparrow$ electrons and $K'$-valley spin-$\downarrow$ electrons, resulting in time-reversed pairs of Chern bands~\cite{Kane2005}, at partial band fillings, the Coulomb interaction fully lifts the spin/valley degeneracy~\cite{crepel2023anomalous}. 
Then, spin/valley polarized electrons in a single Chern band naturally form fractional Chern insulators (FCIs)---the analog of fractional quantum Hall states but enabled by the emergent magnetic field, giving rise to the predicted and observed FQAH effect in twisted bilayer semiconductors~\cite{Devakul_2021, Li2021PRR, crepel2023anomalous}.   

Intriguingly, a recent experiment~\cite{xu2025signaturesunconventionalsuperconductivitynear} reported signatures of superconductivity in $t$MoTe$_2$ emerging at fillings close to $\nu=2/3$ where the FCI state was observed. The coexistence of superconductivity and quantized Hall effect in the same material is remarkable and contrasts with the absence of superconductivity in Landau levels. This surprising  finding calls for theoretical understanding. 

Indeed, we have demonstrated via numerically exact calculation and analytical theory~\cite{GuerciAbouelkomsan2025} that superconductivity can arise from purely repulsive interactions in a (completely) flat Chern $|C| = 1$ band that is a minimal generalization of the lowest Landau level. Moreover, based on numerical results across different system sizes, we observed the ``odd-even'' effect~\cite{Fu_2010} establishing that superconducting state is topological and hosts an odd number of chiral Majorana fermion edge modes.  
   
In this work, we elucidate the key properties of superconductivity in twisted TMDs and its relation to FCIs. 
We establish the off-diagonal long-range order and determine the microscopic superconducting order parameter from exact diagonalization. Due to the presence of strong emergent magnetic field, the superconducting state is a vortex lattice  with one vortex per unit cell, as shown in Fig.~\ref{fig:skyrmion+VL}.  
Unlike Abrikosov vortex lattices~\cite{abrikosov1957magnetic}, each vortex has vorticity $2$, corresponding to $h/e$ flux quanta rather than $h/2e$.

We show that this exotic superconducting state, despite its nonperturbative nature arising from repulsive interaction, is smoothly connected to a chiral topological $f$-wave superconductor in the weak pairing regime. 
Such adiabatic continuity enables us to compute its Chern number using the mean-field description of topological superconductivity~\cite{ReadGreen2000, Ando_2015,antonenko2024making}. 
The resulting Chern number has absolute value $|C|=1/2$, corresponding to a single chiral Majorana edge mode, with its sign opposite to that of the Chern insulator at full band filling $\nu = 1$.

We further study the coexistence and the relation of superconductivity and the FCI in twisted TMDs.
We identify two regimes sketched in Fig.~\ref{fig:skyrmion+VL}(c).
For moderate modulations, a fractional Chern insulator at filling $\nu=2/3$ coexists with a superconductor for $2/3<\nu<1$ that develops chiral $f$-wave superconductivity, which gradually weakens as the filling factor approaches $\nu=1$.
As the modulation of the emergent magnetic field further increase, superconductivity extends down to $\nu = 2/3$ (replacing the FCI) and even persists to lower filling factors. At $\nu=2/3$, the transition from the FCI to the superconductor is found to be first order.

Our work highlights a key difference between twisted TMDs and Landau levels, i.e., between systems with emergent and real magnetic fields. While fractional quantum (anomalous) Hall states appear in both settings, superconductivity is found in twisted TMDs but not in the conventional lowest Landau level. As we show, this difference has a fundamental origin. In a two-dimensional electron gas, the presence of Galilean invariance forbids superconductivity under a uniform magnetic field that preserves continuous magnetic translation symmetry. In contrast, the emergent magnetic field in twisted TMDs is spatially modulated, and the broken Galilean invariance allows for superconductivity in the form of a vortex lattice locked to the moir\'e lattice. 

The importance of Galilean invariance in relation to superconductivity under a magnetic field, or its lack thereof, has been missed in the hypothesized scenario of anyon superconductivity~\cite{laughlinanyon}.  
The superconducting vortex lattice state in our microscopic study of twisted TMDs is a topological superconductor hosting a chiral Majorana edge mode, distinct from lattice models of the superconducting state of anyons~\cite{nosov2025anyon,pichler2025microscopicmechanismanyonsuperconductivity} that do not support unpaired Majoranas.

This work is organized as follows: Sec.~\ref{sec:emergentfield} summarizes the mechanism underlying the emergent magnetic field in twisted TMDs.
Secs.~\ref{sec:VTX} and~\ref{sec:Galilean} examine the consequences of the emergent magnetic field on the superconducting state and demonstrate the importance of broken Galilean invariance for realizing zero-resistance superconductivity.
In Sec.~\ref{sec:Model}, we present the microscopic model. 
Sec.~\ref{sec:ODLRO} presents evidence of off-diagonal long-range order from the eigenvalues of the two-particle reduced density matrix obtained from exact diagonalization. 
In Sec.~\ref{sec:VTXrspace}, we examine the properties of the emergent real space vortex lattice.
Sec.~\ref{sec:ChernN} studies the adiabatic evolution of the superconductor under attractive int    eractions with opposite chiralities $f\!\pm\!if$, establishing a mapping from our strongly interacting model to a weak-coupling description.
This enables us to compute the Chern number of the superconducting phase in Sec.~\ref{sec:topological}. 
Sec.~\ref{sec:phasediagram} focuses on the transition from the FCI to the superconducting phase at $2/3$. 
In Sec.~\ref{sec:coexistence}, we examine the coexistence between the superconductor and the FCI. 
We conclude in Sec. \ref{sec:Discussion} by summarizing our findings and connecting our results to ongoing experiments. 
Additional details supporting our theory are provided in Appendices~\ref{app:model}, \ref{app:2rdm}, ~\ref{app:formfactors}, and~\ref{app:ED}.

\section{Emergent magnetic field in twisted TMDs}\label{sec:emergentfield}

We consider the continuum model of twisted TMDs~\cite{FW_PRL_2019} which describes valence band holes around the $K$ and $K'$ valleys experiencing interlayer tunneling and intra-layer moir\'e potentials. 
The Hamiltonian for holes in one valley involves two layers and is given by, 
\begin{equation}
    H_0 = \frac{\p^2}{2 m^*} \sigma_0 + \J(\r) \cdot \boldsymbol{\sigma} + V(\r) \sigma_0,
    \label{eq:continuum}
\end{equation} 
where $\{\sigma_0,\boldsymbol{\sigma}\}$ are the identity and Pauli matrices in the layer pseudospin space. 
The interlayer tunneling and intralayer moir\'e potentials are encoded in a spatially modulated Zeeman field  $\J(\r)$ and a periodic potential $V(\r)$.

To understand how the emergent magnetic field arises in twisted TMDs, we focus on the large $J$-limit of the Hamiltonian \eqref{eq:continuum}, also known as the adiabatic limit~\cite{Paul_2023,MoralesDuran2024,Shi2024,Kolar2024,Li2025,Onishi_2026}. In such a limit, the layer pseudospin is locally aligned with the Zeeman field giving rise to an effective (one component) Hamiltonian: 
\begin{equation}
\label{eq:adiabatic}
    \tilde{H}_0=\frac{(\p-e\A(\r))^2}{2m^*}+\tilde{V}(\r),
\end{equation}
where $\A(\r)$ is a $U(1)$ vector potential describing an \textit{emergent} spatially modulated magnetic $B(\r) = \nabla \cross \A(\r)$. 
Crucially, in this mapping,  the flux enclosed per unit cell is proportional to the skyrmion winding number of the Zeeman field $\J(\r)$. For twisted TMDs, this corresponds to one flux quantum $h/e$ per moir\'e unit cell. 

This model Hamiltonian $\tilde{H}_0$ supports narrow Chern bands over a wide parameter range~\cite{Paul_2023,MoralesDuran2024}. 
It reduces to the standard Landau level Hamiltonian in the limit of a uniform emergent magnetic field $B$ and a constant scalar potential $\tilde{V}$. 
More generally, the magnetic field and the potential are spatially varying with the periodicity of the underlying moir\'e lattice with lattice vectors $\a_{1/2}$. 
In this case, the emergent magnetic field experienced by the electrons takes the form: 
\begin{equation}\label{emergent_Bfield}
    B(\r) = B_0 + \delta B(\br),
\end{equation}
where the constant part of the magnetic field $B_0 ={h/(e|\bm a_1 \times \bm a_2|)}$ is fixed by the condition of one flux quantum $h/e$ per unit cell $(\a_1,\a_2)$, and $\delta  B(\r)$ is periodic and has zero average over the unit cell.

\section{Emergent vortex lattice}\label{sec:VTX}

Regardless of microscopic details, we expect that in the presence of an emergent magnetic field, the superconducting state hosts an array of quantized vortices, such that the order parameter has a net vorticity of $2$ per unit cell as sketched in Fig.~\ref{fig:skyrmion+VL}. 
Here $2$ is the number of superconducting flux quanta per unit cell in twisted TMDs, i.e, $2e B_0 |\bm a_1 \times \bm a_2| / h=2$.

The form of superconducting order parameter is constrained by lattice symmetry.  
Due to the net magnetic flux, the standard translation operators $T_{\R} = e^{\R \cdot \nabla}$ are replaced by the magnetic translation operators 
\begin{equation}\label{magnetictranslation}
M_{\R} = e^{ie\xi_{\R}(\r)/\hbar} e^{\R\cdot\nabla},    
\end{equation} 
where $\xi_{\R}(\r)$  is the  magnetic phase determined by the requirement  
\begin{equation}\label{magneticphase}
    \nabla \xi_{\R}(\r) = \A_0(\r)-\A_0(\r+\R),
\end{equation}  
and $\A_0(\r)$ is the vector potential of the uniform component of the magnetic field ($B_0=\nabla\times \A_0(\r)$). 
Eqs.~\eqref{magnetictranslation} and~\eqref{magneticphase} are dictated by gauge invariance.
Here, $M_{\a_1}$ and $M_{\a_2}$ commute with each other because of the condition of having one flux quantum per unit cell.

Since the Hamiltonian $\tilde H_0$~\eqref{eq:adiabatic} commutes with the magnetic translation operators $[M_{\R},\tilde H_0]=0$, where $\R=n\a_1+m\a_2$ is a lattice vector, the single-particle orbitals $\psi_{\k}(\r)$ can be labeled by the magnetic crystal momentum $\k$, such that $M_{\R} \psi_{\k}(\r) = e^{i \k \cdot \R} \psi_{\k}(\r)$, or equivalently, 
\begin{equation}\label{eq:boundary_condition1p}
    \psi_{\k}(\r+\R)= e^{i\k\cdot\R}e^{-ie\xi_{\R}(\r)/\hbar}\psi_{\k}(\r).
\end{equation}
The latter relation implies that the magnetic orbitals $\psi_{\k}(\r)$ are quasiperiodic and that the net phase accumulated by a single-particle wave function upon encircling a unit cell is $\Phi_e=\oint d\boldsymbol \ell\cdot e\A_0(\boldsymbol \ell)/\hbar=2\pi$.

Similarly, the superconducting order parameter $\Psi_{\rm pair}$, which describes the wavefunction of charge-$2e$ pair, also transforms nontrivially under translations of the center-of-mass (CoM) coordinate $\X=(\r+\r')/2$ by a lattice vector $\R$. 
Specifically, $\Psi_{\rm pair}$ with zero CoM momentum transforms as:
\begin{equation}
    \label{eq:boundary_condition}
 \Psi_{\rm pair}(\r+\R,\r'+\R)=e^{-ie\frac{\xi_{\R}(\r)+\xi_{\R}(\r')}{\hbar}}\Psi_{\rm pair}(\r,\r').
\end{equation}  
Importantly, the magnetic phase is twice the one acquired by the single particle wavefunction $\psi_{\k}(\r)$~\eqref{eq:boundary_condition1p}, as expected since the Cooper pair carries twice the charge of a single particle. 
Correspondingly, the superconducting order parameter acquires a net phase of $4\pi$ upon transporting the center of mass of the pair around a single unit cell which has magnetic flux $h/e$: 
$\Phi_{\rm pair}=\oint d\boldsymbol \ell\cdot 2 e\A_0(\boldsymbol \ell)/\hbar=4\pi$.
More generally, in the presence of emergent magnetic field with one flux per unit cell,  it follows from gauge invariance and magnetic translation symmetry as described above that the superconducting state is a vortex lattice, with a total vorticity $2N_s$ on the torus.  

\section{Superconductivity Requires Broken Galilean invariance}\label{sec:Galilean}

In the following, we present a general argument demonstrating that broken Galilean invariance is a necessary condition for superconductivity in magnetic fields. 
This rules out the possibility of a translationally invariant superconducting state in a uniform magnetic field, but allows for superconducting states in moir\'e lattices that break continuous translation symmetry.    

For a translationally invariant two-dimensional electron system with a parabolic energy dispersion, 
the Hamiltonian in a uniform magnetic field takes the general form $H=\sum_i ({\bm p}_i - e\bm A_i)^2/2m + \sum_{i<j} U({\bm r}_i - {\bm r}_j)$ with $\bm A_i=(B\hat{\bm z} \times \bm r_i )/2$, where  
we assume that the interaction potential $U$ only depends on relative particle coordinates. Note that $H$ can be re-written in terms of the center-of-mass momentum $\bm P=\sum_i \bm p_i$ and center-of-mass coordinate $\bm R_{0}=\sum_i \bm r_i/N$ ($N$ is particle number) and the remaining $N-1$ relative coordinates and momenta: $H = (\bm P-Ne \bm A)^2/2M + ...$ with $M=Nm$ and $\bm A= (B\hat{\bm z} \times \bm R_{0})/2$. 

Importantly, the center-of-mass variables $\bm P$ and $\bm R_0$ do not appear in the interaction term, or couple to other variables \cite{Kohn1961}.  Furthermore, a uniform electric field $\bm E$ only couples to the center-of-mass coordinate $\bm R_0$ through $- (Ne) \bm E\cdot \bm R_0$, while the total charge current is $\bm J = (Ne) \bm P/M$.  
Thus, this center-of-mass degree of freedom behaves as a single {\it free} ``particle'' of mass $M$ and charge $Ne$ under a perpendicular magnetic field $B$ and electric field $\bm E$. 
It follows from the equation of motion that in the stationary state, the center-of-mass particle moves in the transverse direction with a drift velocity of magnitude $v= E/B$. Therefore, the total current is $\bm J = Ne \v = (N e/B) (\bm E\times \hat{\bm z})$. This implies that the DC conductivity is necessarily $\sigma_{xx}=0, \sigma_{xy} = \rho e / B$ ($\rho$ is particle density), in contrast to a superconductor defined by  $\sigma_{xx}\rightarrow \infty, \sigma_{xy}=0$.

This conclusion holds true even in the presence of attractive interaction that leads to electron pairing. In the presence of a uniform magnetic field, charge-$2e$ Cooper pairs are dispersionless, which frustrates condensation \cite{Yang2008,Barlas2011}. Superconductivity can only develop via the formation of a vortex lattice which spontaneously breaks translational symmetry, as in Abrisokov vortex lattice of conventional superconductors under a weak magnetic field. Moreover, a true zero-resistance state is only realized when the vortex lattice is pinned, consistent with our general conclusion that superconductivity in a magnetic field requires broken Galilean invariance.   

In twisted TMDs, however, the emergent magnetic field~\eqref{emergent_Bfield} is spatially inhomogeneous and preserves only discrete magnetic translational symmetry. In this case, the center-of-mass degree of freedoms $\bm P$ and $\bm R_0$ are no longer decoupled from the other degrees of freedom, hence our previous derivation of the DC conductivity does not apply. 
Importantly, even when the single-particle energy dispersion is completely flat, low-energy charged excitations in these systems are generally \textit{dispersive}~\cite{abouelkomsan_particle-hole_2020,Abouelkomsan2023,schleith2025,goncalves2025,Liu2025,Yan2025}. 
Indeed, the emergent Fermi liquid state in twisted MoTe$_2$ at $2/3<\nu<1$~\cite{reddy2023toward} arises from interaction-induced dispersion for {\it holes} and persists even in the limit of flat electron band~\cite{GuerciAbouelkomsan2025}.

\section{Microscopic Model}\label{sec:Model}

As in our recent work~\cite{GuerciAbouelkomsan2025}, we consider an analytically tractable limit of the Hamiltonian \eqref{eq:adiabatic}, also known as the Aharonov–Casher limit~\cite{MoralesDuran2024,Shi2024,shi2025effectsberrycurvatureideal,dong2022diracelectronperiodicmagnetic}, when the periodic potential cancels the zero-point motion of the emergent magnetic field, giving rise to a perfectly flat $|C| = 1$ band satisfying the trace condition~\cite{Roy2014,Parameswaran2013,Tarnopolsky_2019,JiePRL2021,Ledwith2020,Ledwith2023}:  ${\Tr g(\k)} = |\Omega(\k)|$ with $g(\k)$ is the Fubini-Study metric and $\Omega(\k)$ is the Berry curvature~\cite{resta2020}. 
In this limit, the single-particle wavefunction $\psi_{\k}(\r)$ of the lowest flat Chern band is a generalized lowest Landau level wavefunction, with properties summarized in Appendix~\ref{app:model}.  

The microscopic properties of our model depends on the periodic modulations of the emergent magnetic field $\delta B(\r)$,  which controls the quantum geometry of the $|C| = 1$ band. 
For convenience, we parameterize $\delta B(\r)$ through the Kähler potential $K(\r)$ \cite{JiePRL2021}, which is obtained by solving the Poisson equation, $\nabla^2 K(\br) = e\delta B(\br)/\hbar$. 
We choose $K(\r)$ to be a sum of the lowest harmonics associated with a triangular lattice, 
\begin{equation}
\label{eq:Kofr}
K(\r) = -\dfrac{\sqrt{3}}{4\pi} \mathcal K  \sum_{i = 1,2,3}\cos(\b_i \cdot \r),     
\end{equation} 
with amplitude $\mathcal{K}$, and wave vector modulations  $\b_1=4\pi/(\sqrt{3}a)(1/2,\sqrt{3}/2)$, $\b_2=-4\pi/(\sqrt{3}a)(1,0)$ and $\b_3=-\b_1-\b_2$ and $a$ the moir\'e lattice constant. 

In the following, we work in the symmetric gauge:
\begin{equation}\label{eq:symmetric}
\begin{split}
        &\A_0(\r) = B_0( \hat{z} \cross \r)/2, \\
        &e^{i e \xi_{\R}(\r)/\hbar} = \eta_{\R} e^{i \r \cross \R/2 \ell^2_B},
\end{split}\end{equation}
where $\eta_\R = (-1)^{m n + m +n}$ is the signature of the lattice vector $\R = n \a_1 + m \a_2$, taking the value $\eta_{\R}=+1$ when $\R/2$ is itself a lattice vector and $-1$ otherwise, and $\ell_B$ denotes the emergent magnetic length, $\ell_B = \sqrt{\hbar /e B_0}$.

We further assume that the gap to higher bands is sufficiently large and study the ground state of interacting electrons in the lowest flat band by band-projected exact diagonalization. 
We use shortest-range interaction between spin- and valley-polarized fermions: $V(\r)=v_1\nabla^2\delta^{(2)}(\r)$ with $v_1=3 V_1 a^4/ 4 \pi$~\cite{Trugman1985}.
We refer the interested reader to Ref.~\cite{GuerciAbouelkomsan2025}, and Appendix~\ref{app:model}, for comprehensive details about the full Hamiltonian.

In Ref.~\cite{GuerciAbouelkomsan2025}, we presented numerical and analytical findings of superconductivity in this model in the presence of an inhomogeneous magnetic field, as evidenced by a finite pair binding energy and superfluid stiffness. 
It is remarkable that superconductivity emerges from repulsive interactions in a completely flat band.

\begin{figure}
    \centering
    \includegraphics[width=1\linewidth]{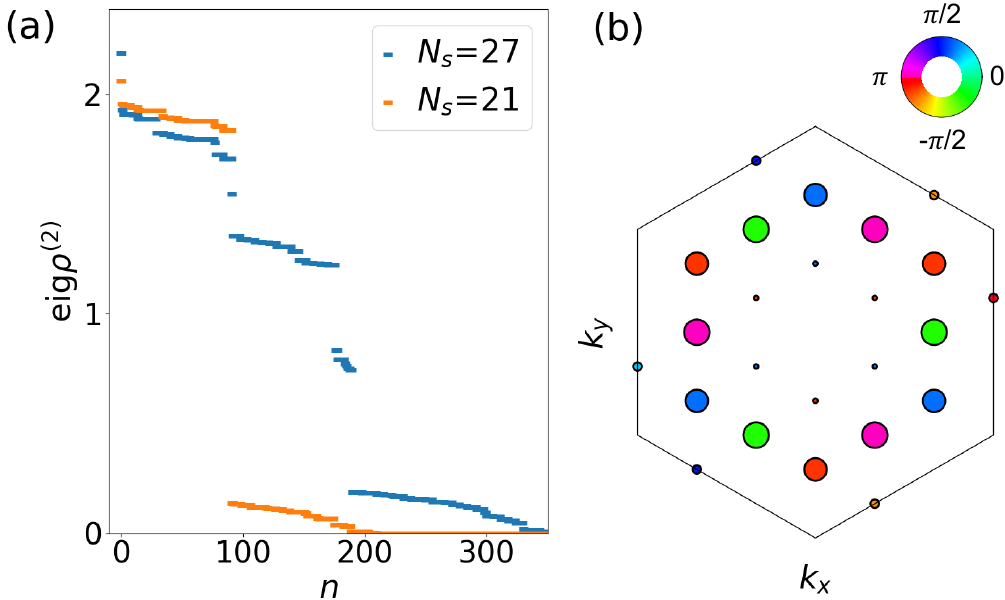}
    \caption{2RDM spectrum  at $2/3$ and condensate wavefunction for $\mathcal K=0.8$: Panel (a) show the spectrum of $\rho^{(2)}$ comparing system sizes $N_s=21,$ 27. 
    Panel (b) shows $\chi_{0(\k,-\k)}$ for $N_s=27$ where the size of the dots quantify the absolute value of $\chi_{0(\k,-\k)}$ while the color its phase.}
    \label{fig:2rdm}
\end{figure}

\section{Off-Diagonal Long-Range Order}\label{sec:ODLRO}

In this and next sections, we elucidate key properties of the superconducting state and reveal the emergent vortex lattice in our model for twisted TMDs (sketched in Fig.~\ref{fig:skyrmion+VL}).

\begin{figure*}
    \centering
    \includegraphics[width=\linewidth]{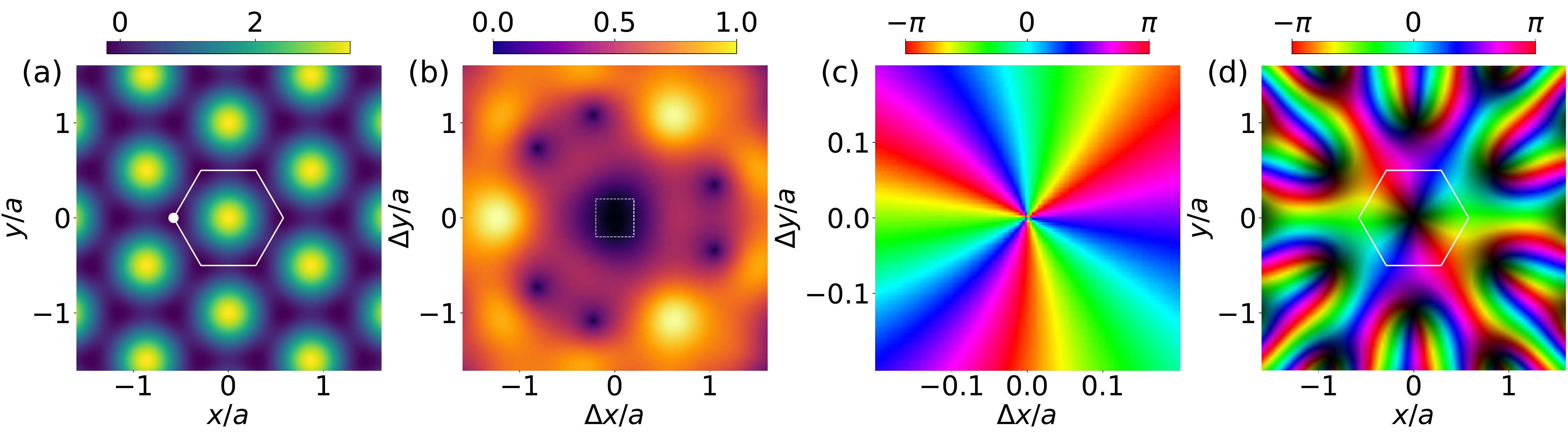}
    \caption{Inhomogeneous magnetic field and the real-space superconducting order: Panel (a) shows the magnetic field $B(\r)$ in unit of $\hbar/(e\ell^2_B)$. 
    Panel (b) shows the absolute value of the pair wavefunction, with the tone (from dark to bright) indicating its magnitude, while panel (c) displays the complex phase at short distances within the square region delineated by the dashed white line in panel (b).
    $\Psi_{\rm pair}$ is shown as a function of the relative coordinate $\Delta \r=\r-\r'$ and $\r'=-a\hat x/\sqrt{3}$. 
    Panel (d) shows the order parameter as a function of the center-of-mass coordinate, where the tone (from black to light) indicates its magnitude and the hue corresponds to the complex phase. 
    Parameters: $\mathcal{K}=0.8$, $N_s=27$, and $\nu=2/3$. 
    White lines in panels (a) and (d) mark the unit cell.}
    \label{fig:psi}
\end{figure*}

First, we corroborate the existence of superconductivity by demonstrating off-diagonal long-range order (ODLRO), which constitutes the defining feature of superconductivity~\cite{RevModPhys.34.694}. 
The pair wavefunction $\Psi_{\rm pair}$, as well as the superconducting fraction, are extracted from the two-particle reduced density matrix (2RDM)~\cite{RevModPhys.34.694}, which we compute from the many-body ground state \(\ket{\Psi}\).
In the magnetic Bloch basis $\psi_{\k}(\r)$, the 2RDM is defined as:
\begin{equation}\label{2rdm}
   \rho^{(2)}_{(\k_1,\k_2)(\k_3,\k_4)}  =\langle c^\dagger_{\k_4}c^\dagger_{\k_3}c_{\k_1}c_{\k_2}\rangle,
\end{equation}
where $\k_{1},\cdots,\k_{4}\in\rm BZ$ and the average is evaluated over the many-body ground state $\ket{\Psi}$, $\langle\cdots\rangle=\mel{\Psi}{\cdots}{\Psi}$.  
This quantity has recently been employed to identify $p+ip$ superconductivity~\cite{li2025attentionneedsolvechiral,li2025berrytrashcanshortrange}.

Introducing $\chi_{n}(\k_1,\k_2)$ as the eigenfunctions of the 2RDM, the operator $\rho^{(2)}$ admits the spectral decomposition:
\begin{equation}
      \rho^{(2)}_{(\k_1,\k_2)(\k_3,\k_4)}=\sum_{n}\lambda_n\chi_{n}(\k_1,\k_2)\chi^*_{n}(\k_3,\k_4).
\end{equation}
Due to momentum conservation, $ \rho^{(2)}_{(\k_1,\k_2)(\k_3,\k_4)}$ is block diagonal in sectors of fixed center-of-mass momentum $\Q = \k_1 + \k_2 \!\!\mod \G$.

Fig.~\ref{fig:2rdm}(a) shows the spectrum of the 2RDM at $\nu=2/3$ and $\mathcal K=0.8$ for $N_{s}=21$ and 27 obtained by performing ED calculations of the Hamiltonian given in Eq.~\eqref{hamiltonian} of Appendix~\ref{app:model}. 
The existence of ODLRO is revealed by the emergence of a large eigenvalue in the center-of-mass (CoM) momentum sector $\Q=0 \mod \G$, which clearly separates from the rest of the spectrum.  
Moreover, the leading eigenvalue $\lambda_0$ increases with system size, $\lambda_0\approx2.06$ for $N_s=21$ and $\lambda_0\approx2.19$ for $N_s=27$. 
The growth of the leading eigenvalue is highly suggestive of ODLRO, where $\lambda_0=\mathcal O(N)$ in the thermodynamic limit~\cite{RevModPhys.34.694}.

We further analyze the eigenfunction $\chi_{0}(\k,-\k)$ associated with the leading eigenvalue $\lambda_0$ of the 2RDM, which corresponds to the pair wavefunction in magnetic Bloch basis.  
Fig.~\ref{fig:2rdm}(b) shows that $\chi_{0}(\k,-\k)$ exhibits the chiral $f-$wave symmetry: it is odd, symmetric  under $C_{3}$ rotation around the center of the unit cell and  behaves as $\chi_{0}(\k,-\k)\propto (k_x-ik_y)^{3}$ for small $\k$, i.e,  transforming as $f-if$. 
This symmetry agrees with the pairing amplitude $\Delta(\k)=\mel{\Psi_{N-2}}{c_{-\k}c_{\k}}{\Psi_N}$ previously obtained from the matrix element of the pair operator between $N-2$ and $N$-particle ground states~\cite{GuerciAbouelkomsan2025}.

Importantly, the emergence of a dominant eigenvalue in $\rho^{(2)}$ is not restricted to $\nu = 2/3$ but persists upon finite doping away from $2/3$, as detailed in Appendix~\ref{app:2rdm}. Together with binding energy and charge stiffness calculations~\cite{GuerciAbouelkomsan2025} (see also Appendix~\ref{app:ED}), these results provide evidence for an extended doping range over which superconductivity emerges.

\section{Real-Space Order Parameter}\label{sec:VTXrspace}

In this section, we determine the real-space superconducting order parameter $\Psi_{\rm pair}(\br, \br')$ by combining the leading eigenfunction $\chi_{0}(\k,-\k)$ of the 2RDM, obtained from band-projected ED, with the single-particle wavefunction $\psi_{\bm k}(\br)$:  
\begin{equation}\label{pair}
    \Psi_{\rm pair}(\r,\r')=\sum_{\k\in \rm BZ}\frac{\chi_{0}(\k,-\k)}{2}\begin{vmatrix}
       \psi_{\k}(\r) & \psi_{\k}(\r')\\
       \psi_{-\k}(\r) & \psi_{-\k}(\r')
    \end{vmatrix},
\end{equation}
where $\begin{vmatrix} ... \end{vmatrix}$ denotes the determinant of the $2\times 2$ matrix. 
Note that $\Psi_{\rm pair}(\r,\r')=-\Psi_{\rm pair}(\r',\r)$ as dictated by fermionic antisymmetry. 

As can also be seen from Eq.~\eqref{pair},  $\Psi_{\rm pair}(\br, \br')$ satisfies the quasi-periodic boundary condition given by Eq.~\eqref{eq:boundary_condition} for a pair of electrons in a magnetic field, because the underlying single-particle wavefunction $\psi_{\bk}(\br)$ satisfies the magnetic translation symmetry described by Eq.~\eqref{eq:boundary_condition1p}. 
In other words, upon moving its center of mass around the unit cell containing $h/e$ flux, generated by the inhomogeneous magnetic field shown in Fig.~\ref{fig:psi}(a), a pair of electrons pick up a phase twice that of a single electron.

The pair wavefunction $\Psi_{\rm pair}$ shown in Fig.~\ref{fig:psi}(b) is obtained by fixing one particle at $\r'=-a\hat{x}/\sqrt{3}$, corresponding to a minimum of the emergent magnetic field (marked by a white dot in Fig.~\ref{fig:psi}(a)), and plotting the wavefunction as a function of the relative coordinate $\Delta\r=\r-\r'$.
Darker tones indicate a small wavefunction absolute value, while brighter regions mark the maxima, revealing three $C_3$-symmetric peaks at a distance $|\Delta \r|/a\approx 1$. 
In Fig.~\ref{fig:psi}(c), the hue represents the complex phase within the white box shown in Fig.~\ref{fig:psi}(b), centered around $\Delta\r=0$.
At short distances, $\Delta\r\to0$, the pair wavefunction exhibits the behavior $\Psi_{\rm pair}(\Delta\r-a\hat x/\sqrt{3},-a\hat x/\sqrt{3})\sim (\Delta x+i\Delta y)^3$, corresponding to a cubic node, consistent with chiral $f$-wave pairing~\cite{GuerciAbouelkomsan2025}.

So far, we have focused on the real-space superconducting order parameter as a function of the relative coordinate, with one of the two particles fixed at a minimum of the emergent magnetic field $B(\r)$~\eqref{emergent_Bfield}; we now turn to its center-of-mass dependence.
Our general argument, presented in Sec.~\ref{sec:VTX}, implies the existence of a vortex lattice featuring two zeros per unit cell as a function of the CoM coordinate, as dictated by the boundary condition in Eq.~\eqref{eq:boundary_condition}.

To reveal the structure of the vortex lattice, we ``integrate out" the relative motion using an \textit{envelope} function that is $C_3$-symmetric and has the appropriate short-distance behavior of the pair wavefunction (see Appendix~\ref{app:vortexlattice}). 
This procedure results in an effective pair wavefunction that depends only on the CoM coordinate. 
As shown in Fig.~\ref{fig:psi}(d), the integrated pair wavefunction features vortices that are pinned to the center of each unit cell.

Remarkably, the pair wavefunction winds \textit{twice} around each vortex, corresponding to a doubled vorticity. 
In addition, these vortices are located in regions where the emergent magnetic field is maximum (Fig.~\ref{fig:psi}(a)). 
This behavior is consistent with the conventional picture that the superfluid density is suppressed where the inhomogeneous magnetic field is strongest.

\section{Adiabatic connectivity}\label{sec:ChernN}

In this section, we show that, despite its strong-coupling origin, the superconducting state induced by repulsive interactions in our model is adiabatically connected to a chiral $f$-wave superconductor arising from attractive interactions.

To this aim, we explore the evolution of the many-body ground state to an attractive interaction with chiral $f$-wave symmetry:
\begin{equation}
\label{pairHamiltonian_main}
     \delta \hat H_m= - g\sum_{\k\p\in \rm BZ}\frac{\mathcal F_m(\k) \mathcal F^*_m(\p)}{N_s}c^\dagger_{\k}c^\dagger_{-\k}c_{-\p}c_{\p},
\end{equation}
where $m = \pm$ labels the two opposite chiralities $f\!\pm\!if$. 
The form factor $\mathcal{F}_m(\k)$ is constructed to induce attractive interactions in the chiral $f$-wave channel, either aligned $(f\!-\!if)$ or anti-aligned $(f\!+\!if)$ with the chirality of the leading eigenfunction $\chi_{0}(\k,-\k)$ of the 2RDM [Fig.~\ref{fig:2rdm}(b)], see Appendix~\ref{app:adiabatic} for details.

\begin{figure}
    \centering
    \includegraphics[width=\linewidth]{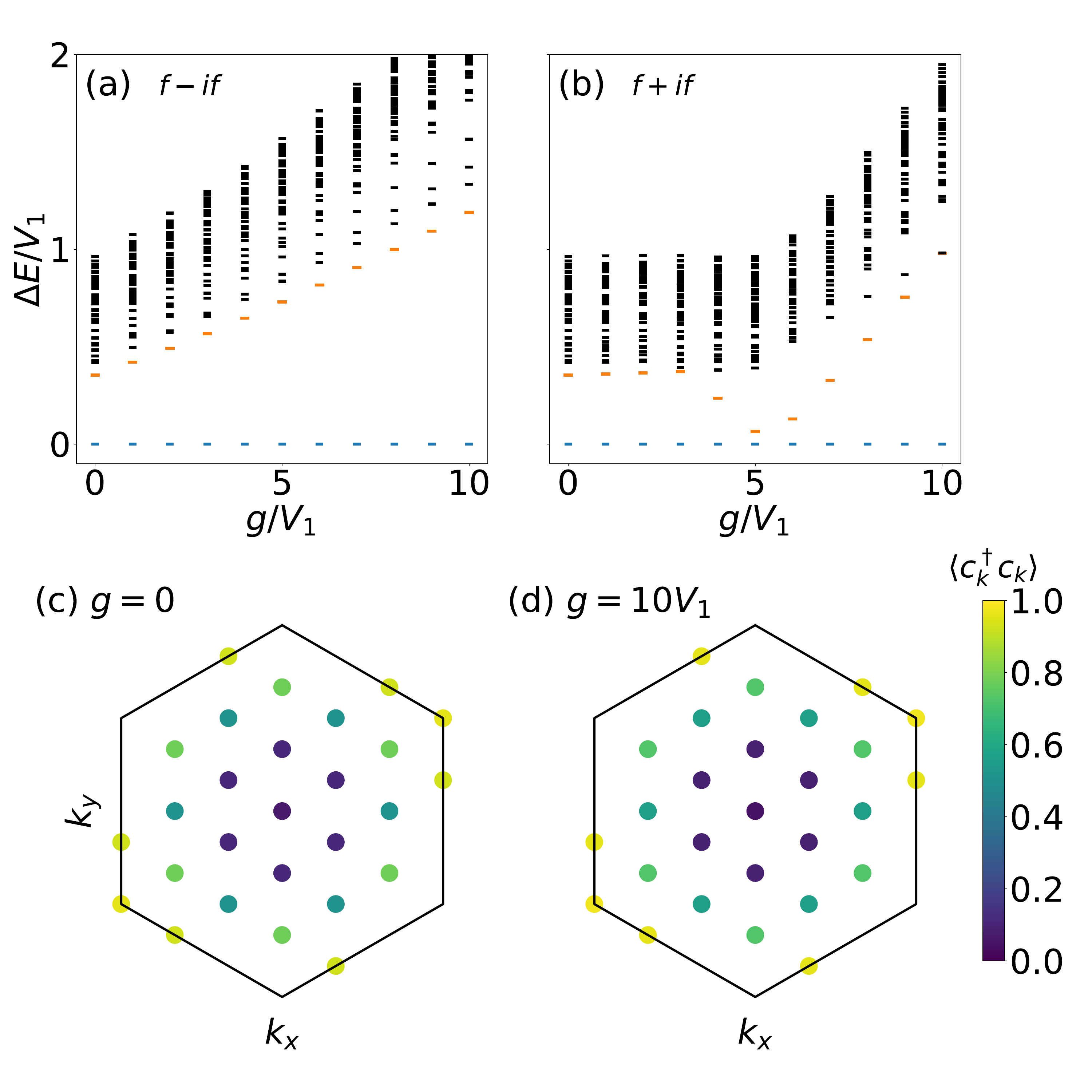}
    \caption{Panels (a) and (b) show the evolution of the many-body energy spectrum as a function of the coupling $g$ for the $f\!-\!if$ ($m=-$) and $f\!+\!if$ ($m=+$) attraction, respectively. Panels (c) and (d) display the single-particle occupation numbers for the unperturbed ground state and for $g/V_1 = 10$ and $f\!-\!if$, respectively.  
    Calculations are performed for $N_s=27$,  $N=16$ and $\mathcal K=0.8$. }
    \label{fig:many_body_spectrum}
\end{figure}

We investigate the effect of adding the pairing term~\eqref{pairHamiltonian_main} to the many-body Hamiltonian presented in Appendix~\ref{app:model} (Eq.~\eqref{hamiltonian})
\begin{equation}
\label{adiabatic_Ham}
\hat  H_m = \hat  H + \delta \hat H_m  .
\end{equation} 
First, we analyze how the unperturbed spectrum ($g=0$) evolves as the coupling $g$ in Eq.~\eqref{pairHamiltonian_main} increases,  keeping $V_1$ and $\mathcal K$ fixed.

The evolution of the many-body spectrum with increasing $g$ reveals two opposite behaviors for $f\!\pm \!if$.
For $f\!-\!if$ ($m=-$), the many-body gap (Fig.~\ref{fig:many_body_spectrum}(a)) grows linearly with the coupling strength, $\delta E_{\rm gap}\sim g$. 
This indicates that preformed Cooper pairs are already present in the $g=0$ ground state and are further stabilized by the chiral $f\!-\!if$ attraction.
In contrast, for the opposite chirality $f\!+\!if$ $(m=+)$, shown in Fig.~\ref{fig:many_body_spectrum}(b), the gap decreases as $g$ increases and a level crossing in the many-body spectrum occurs at $g/V_1\simeq 5$. 
This signals that the pairing interaction $\delta H_{+}$ destroys the superconducting phase and stabilizes a new ground state for $g/V_1>5$. 


The different behavior under $f\!\pm 
\!if$ attractive interaction manifests also in the  behavior of the 2RDM, as shown in Appendix~\ref{app:BdG}. 
We find that the largest eigenvalue $\lambda_0$ increases linearly for $f\!-\!if$ attraction, indicating an enhanced superconducting condensate fraction.
On the other hand, in the $f\!+\!if$ channel, $\lambda_0$ is weakly suppressed as $g$ increases, before exhibiting a sharp jump at $g/V_1 \approx 5$, signaling a quantum phase transition to a superconducting state with opposite chirality.

Taken together, these results confirm that the non-perturbative superconducting state emerging from repulsive interactions in Ref.~\cite{GuerciAbouelkomsan2025} is adiabatically connected to attractive interaction~\eqref{pairHamiltonian_main} with $m=-$, which corresponds to a chiral $f\!-\!if$ superconductor. 
Moreover, both the unperturbed superconducting state and the state realized at $g = 10V_1$ exhibit a single-particle occupation displaying a hole-like Fermi surface, as shown in Figs.~\ref{fig:many_body_spectrum}(c) and~\ref{fig:many_body_spectrum}(d). This behavior is consistent with the interaction-induced dispersion, which has a maximum at $\k = 0$; its explicit form is provided below Eq.~\eqref{phhamiltonian} in Appendix~\ref{app:PHdispersion} (see also Ref.~\cite{GuerciAbouelkomsan2025}). The robustness of the hole-like Fermi surface further supports the topologically nontrivial nature of the superconducting state, placing it in the weak-pairing regime~\cite{ReadGreen2000}, in agreement with the ``odd-even'' effect established in our work~\cite{GuerciAbouelkomsan2025}.

The robustness of the many-body gap and the monotonic enhancement of the leading 2RDM eigenvalue under a chiral $f\!-\!if$ attraction establish an adiabatic path linking the non-perturbative regime to an attractive-interaction description. 
This connection allows us, in the next section, to smoothly deform the system into a weak-coupling mean-field limit, where the topological properties of the superconducting state can be unambiguously characterized by computing its Chern number.

\section{Topological invariant}
\label{sec:topological}

Having established an adiabatic connection to a chiral $f\!-\!if$ superconductor, we now introduce a mean-field Hamiltonian describing the weak-coupling limit.

The corresponding Bogoliubov-de Gennes (BdG) Hamiltonian belongs to the symmetry class D~\cite{Ryu_2010} and reads:
\begin{equation}\begin{split}\label{BdG}
    &\hat H_{\rm BdG} =\int d^2\r \hat \Psi^\dagger (\r)[\tilde H_{0}(\r)-\mu]\hat \Psi(\r)\\
    &+\frac{1}{2}\int d^2\r d^2\r'\left[ \hat \Psi^\dagger(\r) \Delta(\r,\r')\hat \Psi^\dagger(\r') + h.c.\right],
\end{split}\end{equation}
where $\hat \Psi(\r)$ is the Fermi field operator. 
The BdG Hamiltonian~\eqref{BdG} is supplemented by the self-consistency equations for the pairing gap $\Delta(\r,\r')=g\langle\hat \Psi(\r)\hat \Psi(\r')\rangle$, and for the number of particles $N=\int d^2\r\langle\hat \Psi^\dagger(\r)\hat \Psi(\r)\rangle$, which determines the chemical potential $\mu$. 
Without loss of generality, we adopt a BdG Hamiltonian incorporating chiral $f\!-\!if$ pairing interaction and a dispersion with a maximum at $\k=0$, which suffice to determine the topology of our superconducting phase. Additional details are provided in Appendix~\ref{app:BdG}.

\begin{figure}
    \centering
    \includegraphics[width=\linewidth]{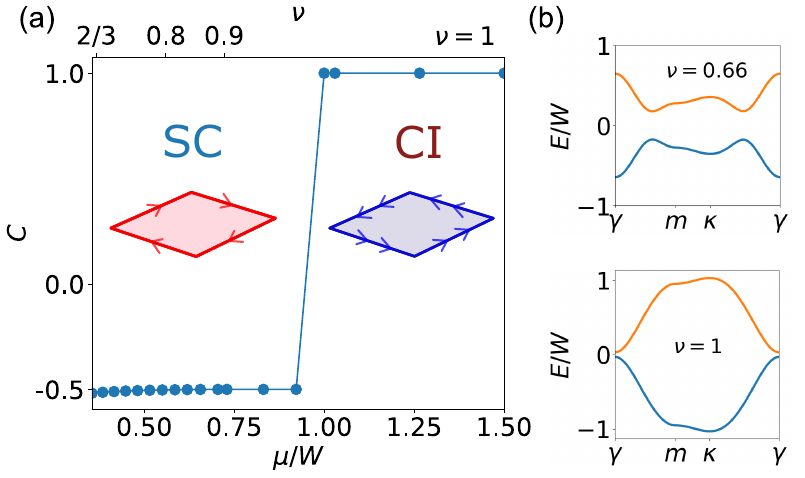}
    \caption{Chern number and BdG bands. Panel (a): Chern number as a function of the chemical potential (lower axis) and filling factor (upper axis). 
    Panel (b): Representative low-energy BdG band structures of the chiral topological superconductor (SC) with $C=-1/2$ $(\nu=0.66)$ and of the Chern insulator (CI) with $C=+1$ $(\nu=1.0)$. 
    $W$ is the bandwidth of the lowest Chern band. }
    \label{fig:chern}
\end{figure}

Fig.~\ref{fig:chern}(a) shows the Chern number of the superconducting ground state as a function of the chemical potential, while Fig.~\ref{fig:chern}(b) illustrates  representative band structures. 
At filling $\nu=1$, where the lowest Chern band is fully filled, the ground state is a Chern insulator (CI) with Chern number $C=+1$, supporting a single chiral complex-fermion edge mode propagating counterclockwise.

Lowering the density, at partial filling of the lowest Chern band, we observe a transition to $ C = -1/2$ corresponding to a single Majorana zero mode with opposite chirality with respect to the Chern insulator at $\nu=1$.  
The $C=-1/2$ state extends down to $\nu = 2/3$, where our exact diagonalization calculations reveal a superconducting vortex lattice.

The topological transition can be understood by decomposing the total Chern number into an orbital contribution inherited from the underlying Chern band and an internal contribution associated with the BdG quasiparticle wavefunctions (see Appendix~\ref{app:BdG}). 
The latter captures the Berry phase from the winding of the BdG wavefunction around the hole-like Fermi surface, shown in Figs.~\ref{fig:many_body_spectrum}(c)-(d), induced by the chiral $f\!-\!if$ gap, yielding a contribution of $-3/2$, while the orbital term contributes $+1$.
Their sum gives a total Chern number \(C=-1/2\), highlighting the crucial role of the underlying band topology. 
By contrast, the strong-pairing limit is topologically equivalent to a Chern insulator with $C=+1$. 
This topological transition is characterized by a change in the number of chiral Majorana edge modes by three.

Furthermore, our results are consistent with the observation in Refs.~\cite{chaudhary2020vortex,Zaletel2019,Schirmer_2022,Shaffer2021hofstadter,antonenko2024making,antonenko2025unifiedtopologicalphasediagram,Liu2015,Tesanovic2000} that vortex lattices with an odd number of superconducting flux quanta per unit cell necessarily correspond to topological superconductors hosting an even number of Majorana modes—unlike the situation realized here, where we have an even number of superconducting flux quanta.

In conclusion, exploiting the adiabatic connection between the $g=0$ ground state and the superconducting state induced by chiral $f\!-\!if$ attraction, we show that the vortex lattice superconductor realizes a topological phase with half-integer Chern number $C=-1/2$, opposite in chirality to the parent Chern insulator.

\begin{figure}
    \centering
    \includegraphics[width=0.95\linewidth]{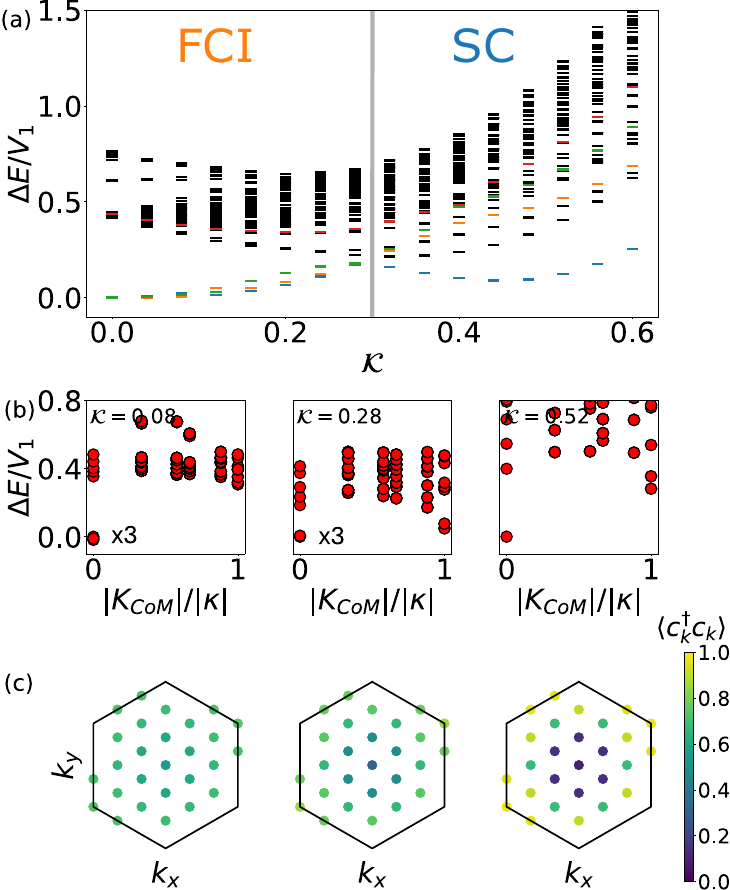}
    \caption{Evolution of the many-body spectrum at $\nu=2/3$: Panel (a) shows the evolution of the many-body spectrum including all momentum sectors as a function of $\mathcal{K}$. 
    The vertical line separates the fractional Chern insulator (FCI) phase from the superconducting (SC) regime. 
    Panels (b) and (c) show the many-body spectrum and the single particle occupation for $\mathcal K=0.08,0.28$ and $0.52$, respectively. 
    Computations are performed for $N_s=27$ unit cells.}
    \label{fig:2A}
\end{figure}

\section{Phase diagram at $\nu = 2/3$}
\label{sec:phasediagram}

Beyond superconductivity, our model features fractional Chern insulators at $\nu = 2/3$ for small values of $\mathcal K$ approaching the uniform field limit ($\mathcal K = 0$) where the Laughlin wavefunction is the exact ground state~\cite{Trugman1985,Ledwith2020,JiePRL2021}. 
Having established superconductivity for large values of $\mathcal K$, we now investigate the interplay of the two phases as $\mathcal K$ is varied.

Fig.~\ref{fig:2A}(a) shows the evolution of the many-body spectrum as a function of $\mathcal K$. 
For $\mathcal K\lesssim0.3$, the spectrum features three low-energy ground states located at the center of mass momentum sector $K_{\rm CoM} = 0$. 
These states are adiabatically connected to the spectrum in the lowest Landau level limit ($\mathcal K=0$), implying that they describe an FCI with threefold degeneracy on the torus.

On the other hand for $\mathcal K >0.3$, we find a unique ground state with $K_{\rm CoM} = 0$.  
To determine the nature of this ground state, we analyze in Fig.~\ref{fig:2A}(c) the single-particle occupation  $\langle c^\dagger_{\k}c_{\k}\rangle$ which shows the formation of a sharp hole-like Fermi surface as $\mathcal K$ increases. 
This behavior signals a transition to a Fermi-liquid–like phase composed of hole-like quasiparticles forming a Fermi sea centered around $\k=0$, originating from the interaction-induced dispersion $\epsilon_{\k}$ below Eq.~\eqref{phhamiltonian}.
Furthermore, as indicated in Fig.~\ref{fig:2A}(b), the transition from the FCI to the compressible phase features a collapse of the magneto-roton gap at $\kappa$ and $\kappa'$.
We observe that a similar behavior is also found for $N_s=21$ unit cells (see Appendix~\ref{app:ED}), where the FCI grounds states carry different CoM momenta.

Indeed, this compressible state corresponds to our superconducting vortex lattice.
As detailed in Appendix~\ref{app:binding}, hole-like quasiparticles near the interaction-induced Fermi surface experience an effective attraction, resulting in a finite binding energy.
Consequently, the Fermi surface becomes unstable to pairing over a broad range of $\mathcal{K}$, leading to the emergence of off-diagonal long-range order and the onset of superconductivity.

To characterize the transition from the FCI to the superconductor beyond single-particle correlations, we compute the \textit{full} structure factor: 
\begin{equation}\label{Sq}
 S(\q)=(\langle\rho(\q)\rho(-\q)\rangle-\langle\rho(\q)\rangle\langle\rho(-\q)\rangle)/A,   
\end{equation}
and analyze its small-momentum behavior. 
In Eq.~\eqref{Sq}, $\rho(\q)=\sum_{\k}f^\dagger_{\k}f_{\k+\q}$ is the unprojected density operator, where $f^\dagger_{\k}$ creates a fermion in a plane wave state with momentum $\k$; here, both $\k$ and $\q$ are unrestricted.

In the FCI, the structure factor exhibits the characteristic form when $\q\to0$:
\begin{equation}
\label{eq:quantumweight}
S(\q) = \frac{K q^2}{4\pi}, 
\end{equation}
where $K$ is the so-called quantum weight~\cite{reddy2023toward,Onishi2024,Onishi2025,Zaklama2025}, satisfying $K \ge |C|$, with $C$ the many-body Chern number — a bound that is saturated in the Landau level limit $\mathcal{K} = 0$. 
In contrast, a superconducting state displays a linear-in-momentum behavior for $\q\to0$, $S(\q) =  \hbar q \rho/(2mc)$, with $m$ the effective mass of the quasiparticles, $\rho=N_h/A$  the density of holes and $c$ the Goldstone mode velocity, following from the Feynman-Bijl relation~\cite{feynmansuperfluid}.

\begin{figure}
    \centering
    \includegraphics[width=.8\linewidth]{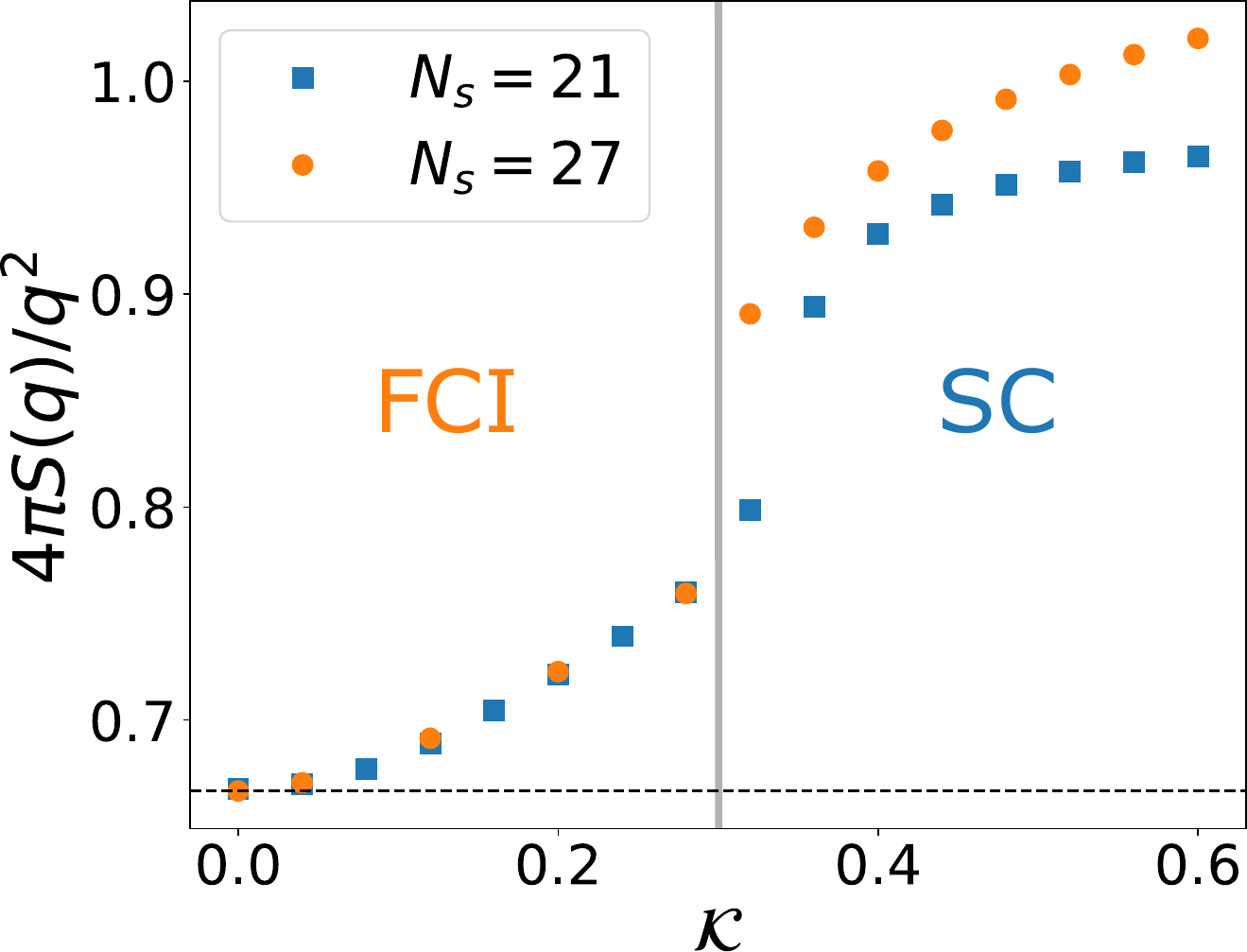}
    \caption{$4\pi S(\q)/q^2$ evaluated at $\q=\Delta \k$ as a function of $\mathcal K$ for $N_s=21$ and $N_s=27$ at $\nu=2/3$.
    The vertical line separates the fractional Chern insulator (FCI) from the superconducting (SC) regime. }
    \label{fig:small_q_Sq}
\end{figure}

Fig.~\ref{fig:small_q_Sq} shows the coefficient $4\pi S(\q)/q^{2}$, evaluated at the smallest nonzero momentum $\q=\Delta \k$ of the finite torus, which sets the momentum resolution. 
In the FCI regime ($\mathcal{K} \lesssim  0.3$), this quantity is finite and converges already for small system sizes ($N_s=21$).
Moreover, it corresponds to the quantum weight \eqref{eq:quantumweight}, with a lower bound set by the absolute value of the many-body Chern number $|C| = 2/3$.

On the other hand, in the superconducting regime ($\mathcal{K}\gtrsim0.3$), we find $4\pi S(\q)/q^{2}$ to increase rapidly and grow with system size. This contrasting behavior signals a transition from the $q^{2}$ scaling characteristic of the FCI to the linear-in-$q$ behavior for superconductors with short-range interactions. Indeed, if the structure factor is linear at small $q$, $4\pi S(\q)/q^{2}$ diverges as $1/\Delta k\propto\sqrt{N_s}$, consistent with our observations in Fig.~\ref{fig:small_q_Sq}.

Importantly, the transition between the two regimes indicates a first-order phase transition, as evidenced by the jump in Fig.~\ref{fig:small_q_Sq} whose magnitude increases with system size and the avoided level crossing observed in the many-body spectrum in Fig.~\ref{fig:2A}(a).

\section{Coexistence of superconductivity and FCI}
\label{sec:coexistence}

\begin{figure}
    \centering
    \includegraphics[width=\linewidth]{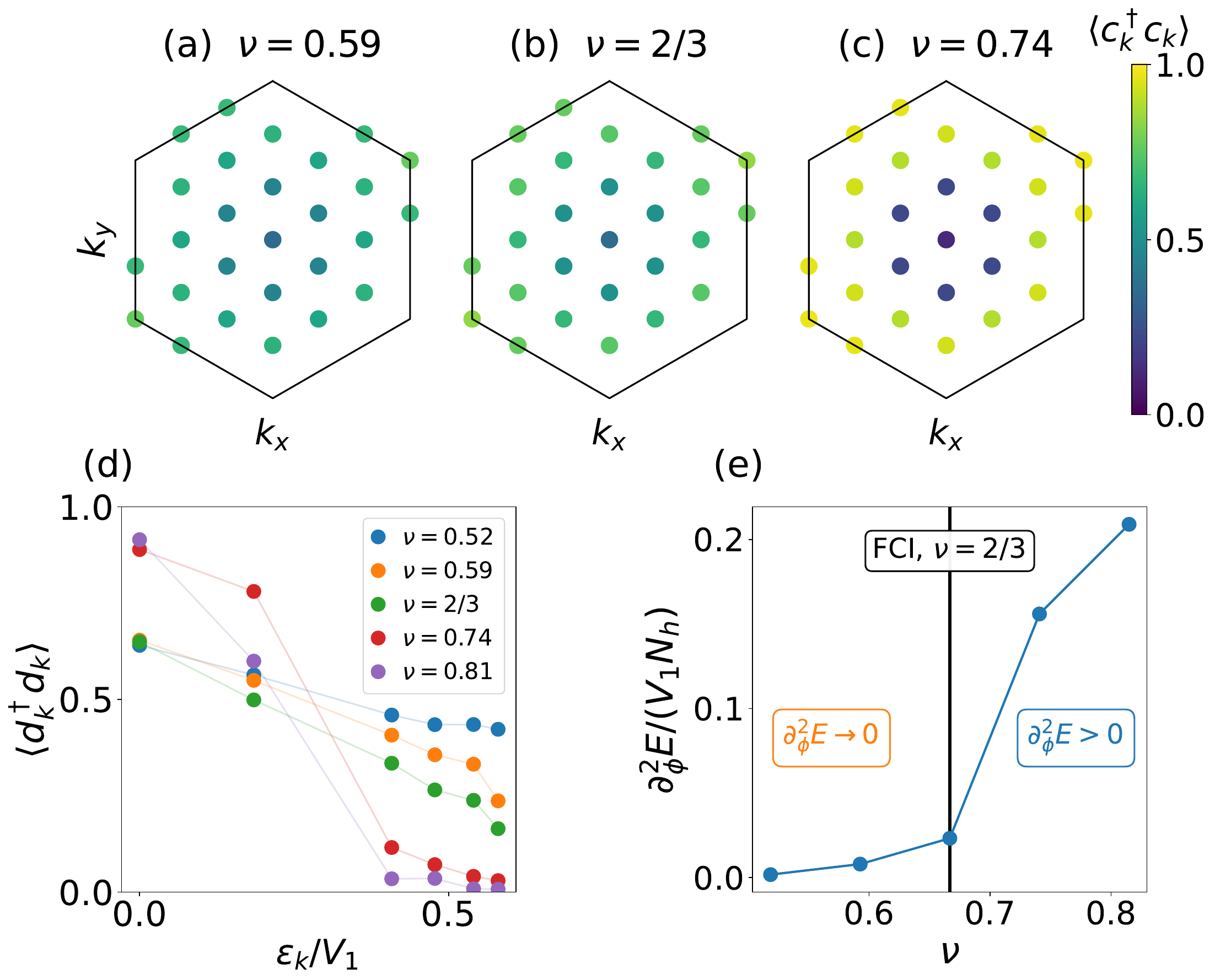}
    \caption{Single particle occupation at $\mathcal{K}=0.25$ in the FCI regime as a function of momentum $\k$ (top panels, (a)-(c)). 
    Panel (d) shows the hole occupation as a function of the interaction-induced dispersion $\epsilon_{\k}$, while panel (e) the phase rigidity evolution across filling $2/3$.}
    \label{fig:n_k_away}
\end{figure}

Having identified the FCI and the superconducting regimes at $\nu = 2/3$ and the onset of the transition between the two as a function of $\mathcal K$, we now show that the superconductor and FCI coexist together in a finite window of $\mathcal K$ at different filling factors. 
To this end, we explore the ground state properties at filling factors away from $2/3$ for $0.2 \lesssim \mathcal K\lesssim 0.3$, where the $\nu=2/3$ is a FCI; see Fig.~\ref{fig:2A}(a). 
Figs.~\ref{fig:n_k_away}(a)-(c) reveals a pronounced particle–hole asymmetry at filling factors away from $2/3$, with qualitatively different behavior above and below $2/3$.

At $\nu = 2/3$ and below, we find no signs of anomalous Fermi surfaces as evident from the single particle occupation $\langle c^\dagger_{\k}c_{\k}\rangle$ which is found to be smooth in momentum space, as shown in Figs.~\ref{fig:n_k_away}(a) and~\ref{fig:n_k_away}(b) for filling factors $\nu=0.59$ and $\nu=2/3$, respectively. 
This is also evident from the hole distribution $\langle d^\dagger_{\k}d_{\k}\rangle$, $\mathcal Pc_{\k}\mathcal P^{-1}=d^\dagger_{-\k}$, as a function of the interaction induced dispersion relation $\epsilon_{\k}$~\eqref{phhamiltonian} shown in Fig.~\ref{fig:n_k_away}(d).

On the other hand, at $\nu > 2/3$, the ground state develops a hole-like Fermi surface, featuring a hole pocket around $\k=0$, in correspondence with the maximum of the interaction induced dispersion relation $\epsilon_{\k}$~\cite{GuerciAbouelkomsan2025}. 
An example is  shown in Fig. \ref{fig:n_k_away}(b) for $\nu = 0.74$. The properties of this ground state are consistent with those of the hole-like superconducting phase observed at $\nu=2/3$ for $\mathcal{K}$ above the critical value; see Fig.~\ref{fig:2A}(a). 


Finally, Fig.~\ref{fig:n_k_away}(e) shows the second derivative of the ground-state energy with respect to the flux $\phi$ threading one handle of the torus, $\partial^2_\phi E/N_h$ with $N_h=N_s-N$ number of holes, quantifying the phase rigidity of the ground state. 
Consistent with the previous analysis, we find two qualitatively different behaviors across the transition. 
For $\nu\le2/3$, the ground state exhibits vanishingly small phase rigidity, implying that the ground state cannot sustain a coherent phase. 
On the other hand, for $\nu>2/3$ the ground state exhibits a finite phase rigidity and can sustain a superconducting state at sufficiently low temperatures~\cite{Scalapino1993}. 

Taken together, these results establish the \textit{coexistence} of superconductivity and FCI in the regime of moderately inhomogeneous emergent magnetic fields.

\section{Discussion}\label{sec:Discussion}

Our work demonstrates the realization of a chiral $f$-wave superconductor hosting an \textit{emergent} moir\'e vortex lattice  in a purely repulsive electron model of twisted TMDs. 
A crucial ingredient for superconductivity is the breaking of Galilean invariance, or continuous magnetic translation symmetry. 
This can be achieved in several ways, including emergent inhomogeneous magnetic fields, as in our model, or externally imposed periodic potentials. 
Notably, a similar chiral $f$-wave superconductor was recently reported in the lowest Landau level with a periodic potential~\cite{wang2025}.  
In both cases, the superconducting state realizes a vortex lattice with pinned vortex cores, enabling a zero-resistance state and exhibiting nontrivial topological properties.

In our case, a key physical ingredient is the interaction-induced dispersion~\cite{abouelkomsan_particle-hole_2020,Abouelkomsan2023} generated from an otherwise flat Chern band. 
This dynamically produces a Fermi surface, from which superconductivity emerges as an instability. 
The robustness of the Fermi surface (see Figs.~\ref{fig:many_body_spectrum}(c)-(d)), together with the sensitivity to boundary conditions, provides conclusive evidence for a weak-pairing topological superconducting phase.

We have presented multiple evidence supporting the existence of superconductivity. 
First, the evaluation of the 2RDM reveals in the superconducting regime a dominant eigenvalue at zero center-of-mass momentum that grows with system size, consistent with off-diagonal long-range order. 
The associated eigenfunction exhibits chiral $f$-wave symmetry and defines the real-space superconducting order parameter, thereby enabling a direct characterization of the vortex lattice in real space.
The superconducting vortex lattice emerges from purely repulsive interactions in an inhomogeneous magnetic field carrying $h/e$ flux quantum per unit cell.
This should be distinguished with other settings where an initially conventional superconductor becomes topological under an external magnetic field that induces a vortex-lattice structure~\cite{Zaletel2019,chaudhary2020vortex,Shaffer2021hofstadter,antonenko2024making,antonenko2025unifiedtopologicalphasediagram,Qi2010chiral,Zocher2016,Kudo2025,Schirmer_2022,Schirmer2024}.

Importantly, the superconducting ground state evolves adiabatically under an attractive interaction with chiral $f$-wave symmetry, implying adiabatic continuity to a chiral topological superconductor. 
This adiabatic connection allows us to determine the topological properties of the ground state by mapping the strongly correlated problem onto an effective weakly interacting description within the Bogoliubov–de Gennes formalism.

The Chern number of the superconducting state with absolute value $|C|=1/2$ reveals a single Majorana zero mode whose chirality opposes that of the Chern insulator at full filling ($\nu = 1$) of the topological flat band. 
This contrasts with recent theoretical works that predict an even number of Majorana zero modes~\cite{shi2024doping,shi2025anyon,nosov2025anyon,Kim_2025,pichler2025microscopicmechanismanyonsuperconductivity}.
Interestingly, in contrast to single vortices in spinless chiral superconductors that can host an isolated Majorana zero mode~\cite{Cheng2009,Biswats2013majoranna,Liu2015,Murray2015majoranna,Gaggioli_2025}, the double vortices found here are expected to have a gapped quasiparticle spectrum in the bulk, consistent with the large pairing gap found in our ED calculation.

Quite generally, our model supports superconductivity over a finite doping window whose extent depends on the modulation strength $\mathcal{K}$. For smaller $\mathcal{K}$, superconductivity appears in the vicinity of the fractional Chern insulator (FCI) at $\nu=2/3$, whereas for larger $\mathcal{K}$ the superconducting region expands to include $\nu=2/3$ and extends to lower fillings. Our numerics (Figs.~\ref{fig:2A} and \ref{fig:small_q_Sq}) are consistent with a first-order transition between the FCI and the superconducting phase as the strength of the quantum geometry is tuned.

Remarkably, our findings are consistent with recent experimental observations in $t$MoTe$_2$ in Ref.~\cite{xu2025signaturesunconventionalsuperconductivitynear} of an anomalous hole-like Fermi Hall liquid for fillings above $\nu = 2/3$ and below $\nu = 1$, which upon cooling evolves into a superconducting state over an extended filling range between $\nu \simeq 0.7$ and $\nu \simeq 0.75$. The superconductor exists in the vicinity of an FCI at $\nu = 2/3$, separated by a pronounced peak in the longitudinal resistivity. This is consistent with our findings at moderate magnetic field modulation, but speaks against the scenario of superconductivity arising from doping anyons to the $\nu=2/3$ FCI.

We predict that the observed superconductor~\cite{xu2025signaturesunconventionalsuperconductivitynear} is a chiral $f$-wave superconductor, which, due to the layer skyrmion winding~\cite{FW_PRL_2019,MoralesDuran2024,Shi2024}, features a strongly spatially modulated superconducting density with $h/e$ vortices and circulating currents on the moir\'e lattice. 
Since the vortices are strongly pinned at the maxima of the emergent magnetic field, they do not induce dissipation; the superconducting state retains zero resistivity for currents below the depinning threshold.
We propose real-space imaging to resolve the spatial modulation of the superconducting density and to detect Majorana zero modes. 

Looking ahead, an interesting future direction is to employ methods that are able to reach larger system sizes \cite{li2025attentionneedsolvechiral}. Another promising direction is to formulate variational wavefunctions that can simultaneously capture both the FCI and the superconductor \cite{abouelkomsan2025topologicalorderdeepstate,abouelkomsan2026first}. 
Our model provides a concrete starting point to microscopically investigate the proximity effect of fractionalization and superconductivity~\cite{clarke2013exotic} in a realistic setting. 

To conclude, our work demonstrates that the spatial modulation of the emergent magnetic field enables and controls the coexistence of superconductivity and fractionalization in twisted TMDs. 
A uniform field induces fractional quantum anomalous Hall states, whereas a superconducting vortex lattice arises when the emergent field forms a periodic array of $h/e$ flux tubes, with each vortex carrying twice the vorticity of Abrikosov vortex. This establishes a unified setting and common mechanism for fractional quantum Hall physics and topological superconductivity---phases traditionally regarded as incompatible.

\section{Acknowledgements}

It is a pleasure to acknowledge helpful discussions with Leonid Glazman, Daniil Antonenko, Inti Sodemann, Aidan Reddy and Filippo Gaggioli. 
This material is based upon work supported by the Air Force Office of Scientific Research under award number FA2386-24-1-4043.
The authors acknowledge the MIT SuperCloud and Lincoln Laboratory Supercomputing Center for providing computing resources that have contributed to the research results reported within this paper. 
A. A. was supported in part by the Knut and Alice Wallenberg Foundation (KAW 2022.0348). 
L. F. was supported in part by a Simons Investigator Award from the Simons Foundation. 

\setcounter{figure}{0}
\setcounter{section}{0}
\setcounter{equation}{0}

\renewcommand{\thefigure}{S\arabic{figure}}

\appendix

\section{Single particle wavefunction and many-body Hamiltonian}\label{app:model}

Our model is obtained by projecting the two-body interaction onto the basis of ideal Chern wavefunctions~\cite{JiePRL2021}, which, in the symmetric gauge, $\A_0(\r)=B_0(\hat{\boldsymbol{z}}\times \r)/2$, take the form 
\begin{equation}\label{app:wavefunction}
 \psi_{\k}(\r)=\mathcal N_{\k}\, e^{-K(\r)}\, \Phi_{\k}(\r),   
\end{equation}
where $\mathcal N_{\k}$ is a normalization factor, $K(\r)$ the Kähler potential and $\Phi_{\k}(\r)$ the lowest Landau level on a torus. 
These wavefunctions arise as zero-mode solutions of the adiabatic Hamiltonian~\eqref{eq:adiabatic} in the so-called Aharonov–Casher limit~\cite{Shi2024}, where the problem reduces to that of a Dirac particle in a spatially modulated magnetic field~\cite{Aharonov_1979,Dubrovin_1980,dong2022diracelectronperiodicmagnetic}.
We consider a magnetic field $B(\r)$ with maxima on a triangular lattice, as shown in the main text. 

The wavefunction satisfies the $\k$-space boundary conditions: $\psi_{\k+\G}(\r)=\eta_{\G}e^{i\ell^2_B(\G\times\k)/2}\psi_{\k}(\r),$ 
where $\eta_{\G}$ is the signature of $\G$ (i.e. 1 if $\G/2$ is a reciprocal lattice vector and $-1$ otherwise). 
Similarly, in real space we have:  $\psi_{\k}(\r+\R)=\eta_{\R}e^{i\k\cdot\R}e^{i\frac{\R\times \r}{2\ell^2_B}}\psi_{\k}(\r),$ where $\R=n\a_1+m\a_2$ are lattice vectors and $\eta_{\R}$ the signature of $\R$ defined below Eq.~\eqref{eq:symmetric}. 
 Moreover, we have employed the symmetric gauge, where the lowest Landau wavefunction $\Phi_{\k}(\r)$ reads: 
\begin{equation}\begin{split}\label{app:lowestLL}
    \Phi_{\k}(\r) &=    \vartheta_1\left(\frac{z}{t_1} -\frac{ k}{G_2},\omega\right)  e^{i\k\cdot\bm t_1\, \frac {z}{t_1}}\\
    &e^{\frac{\pi}{2\Im\omega}\left(\frac{z^2}{t^2_1}-\left|\frac{z}{t_1}\right|^2\right)}e^{\frac{\pi}{2\Im\omega}\left(\frac{k^2}{G^2_2}-\left|\frac{\tilde k}{ G_2}\right|^2\right)},
\end{split}\end{equation}
where $z=x+iy$, $k=k_x+ik_y$, and in complex notation $t_1=-ia$, $G_2=4\pi/\sqrt{3}a$, $t_2/t_1=\omega$, $\omega=e^{2i\pi/3}$, $G_2=it_1/\ell^2_B$ and $G_1=-\omega G_2$. In Eq.~\eqref{app:lowestLL}, we have introduced the Jacobi $\vartheta_1$ function: 
\begin{equation}\label{app:theta_1}
    \vartheta_1( z , \omega) = \sum_{n \in \mathbb{Z} } e^{i \pi \omega (n+1/2)^2} e^{2 i \pi (z-1/2)(n+1/2)}.
\end{equation}
Notice that the single-particle wavefunction satisfies the $C_6$ symmetry relation $\psi_{C_6\k}(C_{6}\r)=e^{i\pi/3}\psi_{\k}(\r)$.

\begin{figure}
    \centering
    \includegraphics[width=\linewidth]{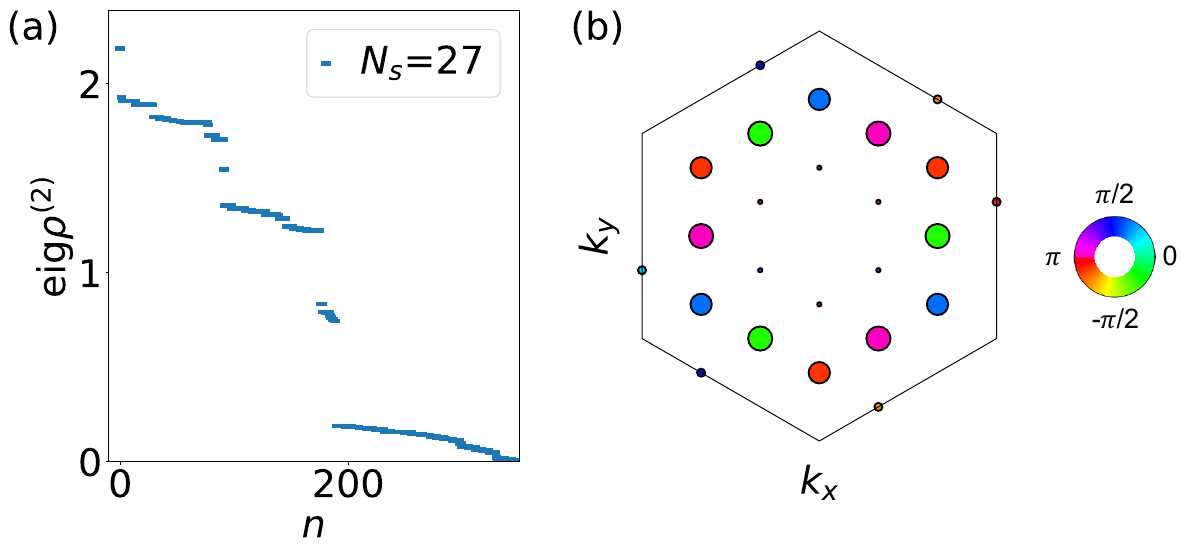}
    \caption{2RDM spectrum (a) and leading eigenfunction (b) for $\mathcal K=0.8$, system size $N_s=27$ and $N=16$ particles.}
    \label{fig:2RDM_16particle}
\end{figure}

Due to the complete flatness of the Chern band, the projected Hamiltonian consists only of two-body $V(\q)$ interactions:
\begin{equation}\label{hamiltonian}
    \hat H=\frac{1}{2A}\sum_{\q}V({\q}):\bar\rho({\q})\bar\rho({-\q}):,
\end{equation}
where $\bar\rho({\q})=\sum_{\k\in \rm BZ}\braket{u_\k}{u_{\k+\q}}c^\dagger_{\k}c_{\k+\q}$ is the projected density operator with $c^\dagger_\k$ creating a particle at momentum $\bk$ in the flat band with wavefunction $\psi_{\k}$~\eqref{app:wavefunction}, $``:\,:"$ denotes normal ordering with respect to the empty band, and $A=|\L_1\times\L_2|$ is the area of the system spanned by the vectors $(\L_1,\L_2)$. 

Finally, we define the wavefunction overlaps $\Lambda_{\k,\k'+\G}=\braket{u_{\k}}{u_{\k'+\G}}$ appearing in the projected density operator in Eq.~\eqref{hamiltonian}: 
\begin{equation}\label{app:form_factors}
\Lambda_{\k,\k'+\G}=\int_{\rm UC}\frac{d^2\r}{|\a_1\times\a_2|} e^{-i\G\cdot\r} u^*_{\k}(\r)u_{\k'}(\r),
\end{equation}
where $\k,\k'\in\rm BZ$ and $u_{\k}(\r)=e^{-i\k\cdot\r}\psi_{\k}(\r)$.

\section{Spectrum of $\rho^{(2)}$ and condensate wavefunction} 
\label{app:2rdm}
To begin, we recall that, following the definition in the main text, $\rho^{(2)}$ is positive semidefinite and satisfies $\Tr\rho^{(2)} = N(N-1)$, where $N$ denotes the number of particles.

Fig.~\ref{fig:2RDM_16particle}(a) shows the spectrum of the 2RDM for $N_s=27$ and $N=16$ particles, corresponding to two-hole doping away from filling $\nu=2/3$. 
The spectrum exhibits a dominant eigenvalue in the momentum sector $\Q=0$, clearly separated from the rest. 
The associated eigenfunction $\chi_{0}(\k,-\k)$, shown in Fig.~\ref{fig:2RDM_16particle}(b), displays the chiral $f-if$ symmetry.

From the 2RDM, we extract the condensate wavefunction corresponding to the largest eigenvalue. 
This can be readily obtained starting from the definition~\cite{RevModPhys.34.694}: 
\begin{equation}
\rho^{(2)}(\r_1,\r_2;\r'_2,\r'_1)=\langle \hat\Psi^\dagger(\r'_1)\hat\Psi^\dagger(\r'_2)\hat\Psi(\r_1)\hat\Psi(\r_2)\rangle,
\end{equation}
where $\hat\Psi(\r)$ is the Fermi field operator. 
Expanding $\hat \Psi(\r)$ in the flat Cher band basis, we find: 
\begin{widetext}
\begin{equation}\begin{split}
\rho^{(2)}(\r_1,\r_2;\r'_2,\r'_1)&=\sum_{\k_1\cdots\k_4\in \rm BZ}\psi^*_{\k_4}(\r'_1)\psi^*_{\k_3}(\r'_2)\psi_{\k_1}(\r_1)\psi_{\k_2}(\r_2)\langle c_{\k_4}^\dagger c_{\k_3}^\dagger c_{\k_1}c_{\k_2}\rangle\\
&=\sum_{\k_1\cdots\k_4\in \rm BZ}\sum_{n}\lambda_n \psi_{\k_1}(\r_1)\chi_{n(\k_1,\k_2)}\psi_{\k_2}(\r_2)\, \psi^*_{\k_3}(\r'_2)\chi^*_{n(\k_3,\k_4)}\psi^*_{\k_4}(\r'_1)\\
&=\lambda_0\Psi_{\rm pair}(\r_1,\r_2
)\Psi_{\rm pair}(\r_2',\r'_1)^* + \cdots.
\end{split}\end{equation}
\end{widetext}
In the final line, we extract the contribution arising from the leading eigenvector, namely the eigenvector associated with the largest eigenvalue $\lambda_0$. The corresponding wavefunction $\Psi_{\rm pair}$~\eqref{pair} describes the pair wavefunction which is macroscopically occupied in the superconducting ground state.

\begin{figure}
    \centering
    \includegraphics[width=\linewidth]{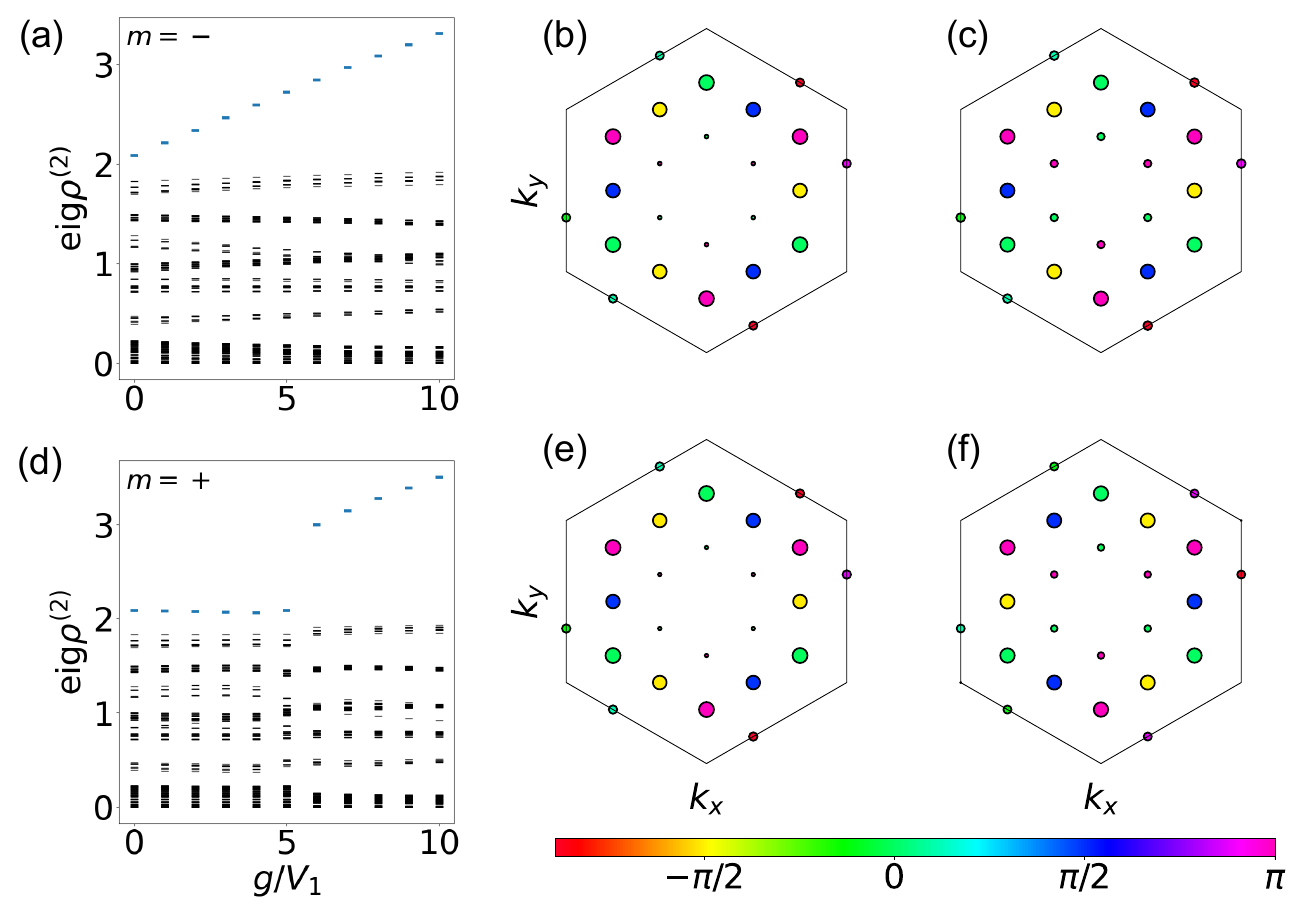}
    \caption{Evolution of the 2RDM spectrum and the Cooper pair wavefunction as a function of $g$ for $m=\pm$: Panels (a) and (d) show the spectra of $\rho^{(2)}$ for $m=-$ and $m=+$, respectively.
    Panels (b) and (c) display the leading eigenvector of $\rho^{(2)}$ for $g/V_1 = 1$ and $g/V_1 = 10$ with chirality $m = -$, respectively. Conversely, panels (e) and (f) show the corresponding leading eigenvectors for the same values of $g/V_1$ but opposite chirality $m = +$.}
    \label{fig:2rdm_adiabatic}
\end{figure}

\subsection{Momentum space properties of the superconducting order parameter}

The leading eigenfunction is odd, symmetric under $C_3$ rotations:
\begin{equation}\label{app:C3symmetry}
    \chi_{0}(C_3\k,-C_3\k)=\chi_{0}(\k,-\k).
\end{equation}

Moreover, in Chern bands, the pair wavefunction carries a momentum-space Berry flux equal to twice that of the underlying single particle wavefunction. 
This implies the momentum-space boundary condition, which, in the symmetric gauge, is given by: 
\begin{equation}\label{app:kBC}
\Delta(\k)=e^{i\ell^2_B(\G\times\k)}\Delta(\k+\G).
\end{equation}
The same boundary condition is satisfied by $\chi_0(\k,-\k)$.

As a consequence of Eq.~\eqref{app:kBC}, there are six unpaired momenta at which the pairing function vanishes, located at $\gamma$, $\kappa$, and $\kappa'$, as well as at the three different $m$ points. 
For $f\!-\!if$ pairing, the gap function exhibits a cubic zero at $\k=0$, corresponding to angular momentum $-3$. In addition, we find simple zeros at $\kappa$ and $\kappa'$ with angular momentum $-1$, and at the three $m$ points with angular momentum $+1$. 
The total vorticity is therefore $-2$, consistent with the $\k$-space Berry flux enclosed by the superconducting gap upon transport along the boundary of the Brillouin zone, Eq.~\eqref{app:kBC}.

\subsection{Real-space properties of the pair wavefunction}\label{app:relative}

The pair wavefunction $\Psi_{\rm pair}(\r,\r')$ depends separately on $\r$ and $\r'$, varying on the moir\'e length scale. We first discuss its dependence on the relative coordinate $\x=\r-\r'$, keeping the center-of-mass coordinate $\X=(\r+\r')/2$ fixed;
the properties of $\X$ are discussed in the next section.

As a function of the relative coordinate $\x$, at short distances ($\x \rightarrow 0$), the wavefunction $\Psi_{\rm pair}$ vanishes as $\Psi_{\rm pair}\sim z^{3}$, where $z$ denotes the holomorphic relative coordinate. 
However, an exception occurs at the special points $\X = n\a_1 + m\a_2$, where the center of mass lies exactly at the double vortices of the vortex lattice
(Fig.~\ref{fig:psi}(d)). 
In such case, the relative wave function vanishes instead  as $z^{5}$. 
This enhanced nodal order arises from the combined effect of the microscopic chiral $f\!-\!if$ pairing symmetry, which contributes a cubic zero, and the double vortices located at $\X = n\a_1 + m\a_2$.

\subsection{Vortex Lattice Structure and Center-of-Mass Dependence}\label{app:vortexlattice}

To extract the dependence on the center of mass, we ``integrate out" the relative coordinate $\x$:
\begin{equation}
    \psi(\X)=\int d^2\x F(\x) \Psi_{\rm pair}(\X+\x/2,\X-\x/2),
\end{equation}
where $F(\x)$ is odd $F(-\x)=-F(\x)$. 
This condition on $F(\x)$ follow from the antisymmetry of the pair wavefunction, $\Psi_{\rm pair}(\X-\x/2,\X+\x/2)=-\Psi_{\rm pair}(\X+\x/2,\X-\x/2)$. 
Moreover, we require that $F(\x)$ reproduce the short-distance $C_3$-symmetric properties of $\Psi_{\rm pair}$.
We readily realize that, in the symmetric gauge~\eqref{eq:symmetric}, $\psi(\X)$ satisfies the magnetic boundary conditions: 
\begin{equation}
    \psi(\X+\R) = e^{i \R\times\X/\ell^2_B}\psi(\X),
\end{equation}
where $\R$ is a lattice vector. 

In the following, we prove that $\psi(\X)$ hosts a vortex lattice of double vortices pinned at $\R=n\a_1+m\a_2$. 
To this aim, we first notice that:
\begin{equation}\begin{split}\label{app:c3pair}
    &\Psi_{\rm pair}(\X+C_3\x/2,\X-C_3\x/2)=\omega^2\\
    &\Psi_{\rm pair}(C^{-1}_3\X+\x/2,C^{-1}_3\X-\x/2),
\end{split}\end{equation}
where in the last step we utilized the $C_{3}$ symmetry in Eq.~\eqref{app:C3symmetry}. 
Furthermore, we have used the three-fold rotational symmetry around $z$ of the wavefunction $\psi_{C_3\k}(C_3\r)=\omega \psi_{\k}(\r)$.

From Eq.~\eqref{app:c3pair}, follows that: 
\begin{equation}\label{app:psiC3}
    \psi(C_{3}\X)=\omega^2\psi(\X).
\end{equation}
Consequently, we have $\psi(0) = \omega^2 \psi(0)$, which immediately implies $\psi(0) = 0$. 
The combination of flux conservation—requiring two superconducting flux quanta per unit cell—and the symmetry relation~\eqref{app:psiC3} implies $\psi(\X \to 0) \sim (X + i Y)^2$. 
As a consequence of discrete magnetic translation symmetry, the double vortices form a vortex lattice.

\section{Form factors of the attractive interaction}\label{app:formfactors}

To establish the adiabatic connection of the chiral $f-if$ superconductor with the weak-pairing regime, we explore the evolution of the many-body ground state under the application of an attractive interaction~\eqref{pairHamiltonian_main}.

The chirality $m=-$, characterized by the form factor $\mathcal{F}_-(\k)$, is constructed to generate two-particle bound states with the same symmetries and boundary conditions of the leading eigenfunction of the 2RDM, namely $\chi_{0(\k,-\k)}$.
Accordingly, $\mathcal F_-(\k)$ is required to satisfy three conditions. First, it must be invariant under $C_{3}$ rotations, $\mathcal F_-(C_{3}\k)=\mathcal F_-(\k)$, and odd under inversion $\mathcal F_-(-\k)=-\mathcal F_-(\k)$. Second, it exhibits small-momentum behavior, $\mathcal F_-(\k)\sim (k_x-ik_y)^3$. Finally, it obeys the momentum-space boundary condition inherited from the flat Chern band wavefunction:
\begin{equation}
\mathcal F_-(\k) = e^{i\ell^2_B(\G\times\k)} \mathcal F_-(\k+\G),
\end{equation}
for any reciprocal lattice vector $\G$.

A function that satisfies these conditions is expressed as follows: 
\begin{widetext}
\begin{equation}\label{attraction_fwave}
 \mathcal F^*_-(\k)= \varphi(\k)\left[\vartheta_1\left(\frac{k}{G_2},\omega\right)e^{\frac{\pi}{2\Im\omega}\left[\left(\frac{k}{G_2}\right)^2-\left|\frac{k}{G_2}\right|\right]}\right]^2,\quad \varphi(\k)=\sum_{\G}e^{-\ell^2_B|\k+\G|^2/2}a(k+G)_+, 
\end{equation}
\end{widetext}
where $\vartheta_1$ is the Jacobi Theta function~\eqref{app:theta_1} and $\varphi (\k)$ is periodic ($\varphi (\k+\G)=\varphi(\k)$) and behaves like $p+ip$ for small $\k$.  
Finally, the form factor with opposite chirality is defined as $\mathcal F_+(\k)=\mathcal F_-^*(\k)$.

\subsection{Adiabatic evolution under $\delta H_m$}\label{app:adiabatic}

We examine the 2RDM spectrum as a function of pairing strength $g$, with $g=0$ as the unperturbed regime. Its evolution depends sensitively on the small-momentum chirality of the form factor, $\mathcal F_\pm(\k) \sim (k_x \pm i  k_y)^3$ for $\k\to0$, leading to qualitatively distinct behaviors. 

For $f\!-\!if$ $(m=-)$, aligned with the chirality of $\chi_{0(\k, -\k)}$ at $g=0$, the many-body gap grows linearly with $g$, and the leading eigenvalue of the 2RDM increases correspondingly, indicating a progressive increase of the superfluid fraction (Fig.~\ref{fig:2rdm_adiabatic}(a)).
Fig.~\ref{fig:2rdm_adiabatic}(b) and Fig.~\ref{fig:2rdm_adiabatic}(c) show the momentum space structure of the eigenfunction $\chi_{0(\k,-\k)}$ obtained for $g/V_1=1$ and for $g/V_1=10$ and chirality $f\!-\!if$. 
Increasing the coupling strength $g$ does not change the $f\!-\!if$ character of the leading eigenfunction; however, it enhances the absolute value of $\chi_{0(\k,-\k)}$ away from the Fermi surface. 

Conversely, for $f\!+\!if$ $(m=+)$, with chirality opposite to that of $\chi_{0(\k,-\k)}$ at $g=0$, the leading eigenvalue shown in Fig.~\ref{fig:2rdm_adiabatic}(d) initially decreases slightly as $g$ increases, then exhibits a sharp jump around $g/V_1\sim 5$, coinciding with the level crossing in the many-body ground state. Increasing $g$ from $g/V_1=1$ (Fig.~\ref{fig:2rdm_adiabatic}(e)) to $g/V_1=10$ (Fig.~\ref{fig:2rdm_adiabatic}(f)) drives a change in the chirality of the leading eigenvector from $f\!-\!if$ to $f\!+\!if$. This is consistent with a quantum phase transition between two superconducting states of opposite chirality. 

\subsection{BdG formalism}\label{app:BdG}

\begin{figure}
    \centering
    \includegraphics[width=\linewidth]{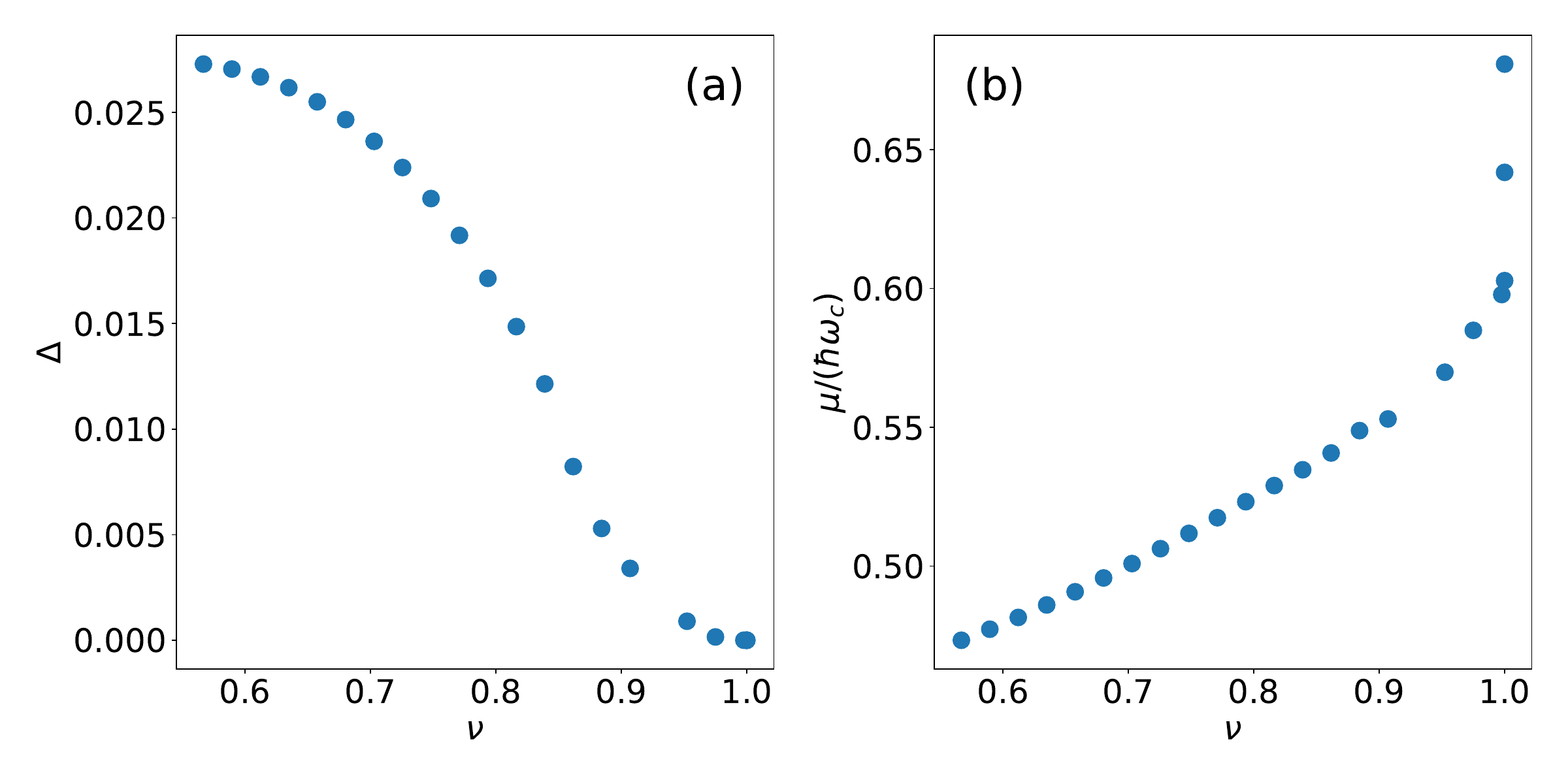}
    \caption{Panels (a) and (b) show the pairing gap and the chemical potential at self-consistency as a function of the filling factor $\nu$. Calculations are performed setting $v=V/(\hbar \omega_c)=-0.2$ and $g/(\hbar\omega_c)=1.5$ for $N_s=21\times 21$ unit cells.}
    \label{fig:sc_BdG}
\end{figure}

Exploiting the adiabatic connectivity of our model to electrons with $f\!-\!if$ attraction, we compute the Chern number of the superconducting state within the Bogoliubov-de Gennes (BdG) framework. 
Without losing generality, $\tilde{H}_0(\r)$ [Eq.~\eqref{eq:adiabatic}] realizes one flux quantum per unit cell and an interaction-induced dispersion maximized at $\k=0$~\cite{GuerciAbouelkomsan2025}. The simplest single-particle Hamiltonian satisfying these requirements reads
\begin{equation}\label{app:adiabatic1}
    \tilde{H}_0 = \frac{(\p-e\A(\r))^2}{2m^*}+V\sum_{j=1}^3\cos(\b_j\cdot\r),
\end{equation}
with $V<0$ and uniform magnetic field $B_0=\nabla\times\A(\r)$.

For the interacting part, we assume that pairing gap $\Delta(\r,\r')$ develops exclusively within the lowest Chern band, corresponding to an effective attraction smaller than the band gap~\cite{Akera1991,MacDonald1992,Tesanovic1992}. 
This assumption is motivated by our band-projected ED results, which show that superconductivity develops entirely within a single Chern band. 
We further take $\Delta(\r,\r')$ to  arise from the attractive interaction with strength $g$ in Eq.~\eqref{pairHamiltonian_main} with $f\!-\!if$ symmetry, for which we have previously established adiabatic continuity. 
The resulting BdG Hamiltonian~\eqref{BdG} is solved self-consistently, as detailed below.

We begin by projecting $\tilde H_0$~\eqref{app:adiabatic1} onto the Landau-level basis ${\ket{\phi_{\k n}}}$. 
Measuring energies in unit of $\hbar\omega_c$, we obtain:
\begin{equation}\label{app:LLprojected}
   \tilde H_{0}(\k) = \mathbf 1\left(n+\frac{1}{2}\right) + v\sum_{j=1}^{3}\frac{I(\k,\b_j)+I(\k,-\b_j)}{2},
\end{equation}
where $v=V/\hbar\omega_c$ where the energy is expressed in unit of $\hbar\omega_c$, and $H_0(\k)$ is a matrix in the Landau level index $n$.
The matrix elements $I_{nn'}(\k,\G)$ is obtained employing magnetic translation algebra: 
\begin{widetext}
\begin{equation}
I_{nn'}(\k,\G)=\mel{\phi_{\k n}}{e^{-i\G\cdot\r}}{\phi_{\k n'}}=\eta_{\G}e^{i\ell^2_B(\k\times\G)}e^{-\ell^2_B|\G|^2/4} J_{nn'}\left(\frac{\ell_B G_+}{\sqrt{2}},\frac{\ell_B G_-}{\sqrt{2}}\right),
\end{equation}
where 
\begin{equation}
     \begin{split}
     &J_{nn'}\left(\frac{\ell_B G_+}{\sqrt{2}},\frac{\ell_B G_-}{\sqrt{2}}\right)=\sqrt{\frac{n!}{n'!}}\left(-\frac{\ell_B G_-}{\sqrt{2}}\right)^{n'-n}L_n^{(n'-n)}\left(\frac{\ell^2_B|\G|^2}{2}\right),\quad \text{if}\, n'\ge n,\\
    &J_{nn'}\left(\frac{\ell_B G_+}{\sqrt{2}},\frac{\ell_B G_-}{\sqrt{2}}\right)=\sqrt{\frac{n'!}{n!}}\left(\frac{\ell_B G_+}{\sqrt{2}}\right)^{n-n'}L_{n'}^{(n-n')}\left(\frac{\ell^2_B|\G|^2}{2}\right),\quad \text{if}\, n> n',
    \end{split}     
\end{equation}
\end{widetext}
where $L^{(n)}_m(x)$ is the generalized Laguerre polynomials.
We emphasize that the size of the matrix in Eq.~\eqref{app:LLprojected} scales with the number of Landau levels retained in the calculation, $N_{\rm LL}$. Since we are interested only in the lowest Landau level band, we set $N_{\rm LL}=5$ in our calculations.

Within the mean-field approximation, in the basis $\Phi_{\k}=(c_{\k0},\cdots,\, c^\dagger_{-\k0},\cdots)^T$ with $c_{\k n}$ annihilating a fermion in $n$-th Landau level at momentum $\k$, the BdG Hamiltonian takes the form: 
\begin{equation}\label{eq:BdGkspace}
    \hat H_{\rm BdG}= \frac{1}{2}\sum_{\k\in \rm BZ}\Phi^\dagger_{\k}\begin{pmatrix}
        H_0(\k)-\mu & \Delta(\k) \\ 
        \Delta(\k)^\dagger & -H_0^T(-\k)+\mu
    \end{pmatrix}\Phi_{\k},
\end{equation} 
where we have introduced the chemical potential $\mu$ and the pairing term: 
\begin{equation}
    \Delta(\k)_{nn'} =
   -g\Delta \mathcal F_-(\k)\delta_{nn'}\delta_{n0}.
\end{equation}

In the latter expression, the value of $\mu$ and $\Delta$ are obtained by solving the self-consistency equations: 
 \begin{equation}\label{delta_number}\begin{split}
     &\Delta = \frac{1}{N_s}\sum_{\k\in\rm BZ} \mathcal F^*_-(\k)\langle c_{-\k0} c_{\k0}\rangle,\\
     &N=\sum_{\k\in\rm BZ }\sum_n\langle c^\dagger_{\k n}c_{\k n}\rangle,
\end{split}\end{equation}
 where $N$ is the particle number.  Fig.~\ref{fig:sc_BdG}(a) and~\ref{fig:sc_BdG}(b) shows the evolution $\Delta$ and $\mu$ obtained solving the self-consistency equations~\eqref{delta_number}.
We emphasize that increasing the filling factor $\nu$ suppresses the pairing strength, since the magnitude of the pairing function on the hole-like Fermi surface scales as the cube of the Fermi momentum, $|\mathcal F_-(\k_F)| \propto |k_F|^3$, with $k_F^2$ proportional to the hole filling, $k_F^2 \propto 1-\nu$ ($\nu\le1)$.

To obtain the Chern number of the superconducting ground state, we first recall the eigenstates of the BdG Hamiltonian: 
\begin{equation}
     \ket{ \phi_{\k\alpha}} = \begin{pmatrix}
        U_{0\alpha}(\k) \ket{\phi_{\k 0}},\cdots,V_{0\alpha}(\k) \ket{\phi^*_{-\k n}},\cdots
     \end{pmatrix}^T.
\end{equation}
The Berry connection is given by: 
\begin{equation}\begin{split}
    &\mathcal A_{\alpha}(\k) = -i\braket{\phi_{\k\alpha}}{\partial_{\k}\phi_{\k\alpha}} = \mathcal A_{\rm LL}(\k) Q_{\alpha}(\k)\\
    &-i\sum_n\left[U^*_{n\alpha}(\k)\partial_{\k}U_{n\alpha}(\k)+V^*_{n\alpha}(\k)\partial_{\k}V_{n\alpha}(\k)\right],
\end{split}\end{equation}
where $\mathcal A_{\rm LL}(\k)$ is the Berry connection of Landau levels and we have introduced the quantity: 
\begin{equation}
    Q_{\alpha}(\k)=\sum_{n}\left[|U_{n\alpha}(\k)|^2-|V_{n\alpha}(\k)|^2\right].
\end{equation}
$Q_\alpha(\k)$ is constant away from zero energy, taking the value $+1$ for electron-like bands and $-1$ for hole-like ones. Finally, the BdG Chern number $C$ is given by: 
\begin{equation}\label{app:chern_number}
   2C=\sum_{\alpha}^{E_{\alpha}<0}C_{\alpha}+N_{\rm LL},
\end{equation}
where $N_{\rm LL}$ is the number of Landau levels included in the calculation and the sum extends to the BdG states with negative energy. 
This definition guarantees the Chern number is measured with respect to the vacuum which is topologically trivial. 
Moreover, the factor $2$ in Eq.~\eqref{app:chern_number} reflects that, in our convention, the Chern number counts complex fermions and, therefore, each Majorana mode corresponds to a half-integer contribution.

\begin{figure}
    \centering
\includegraphics[width=0.9\linewidth]{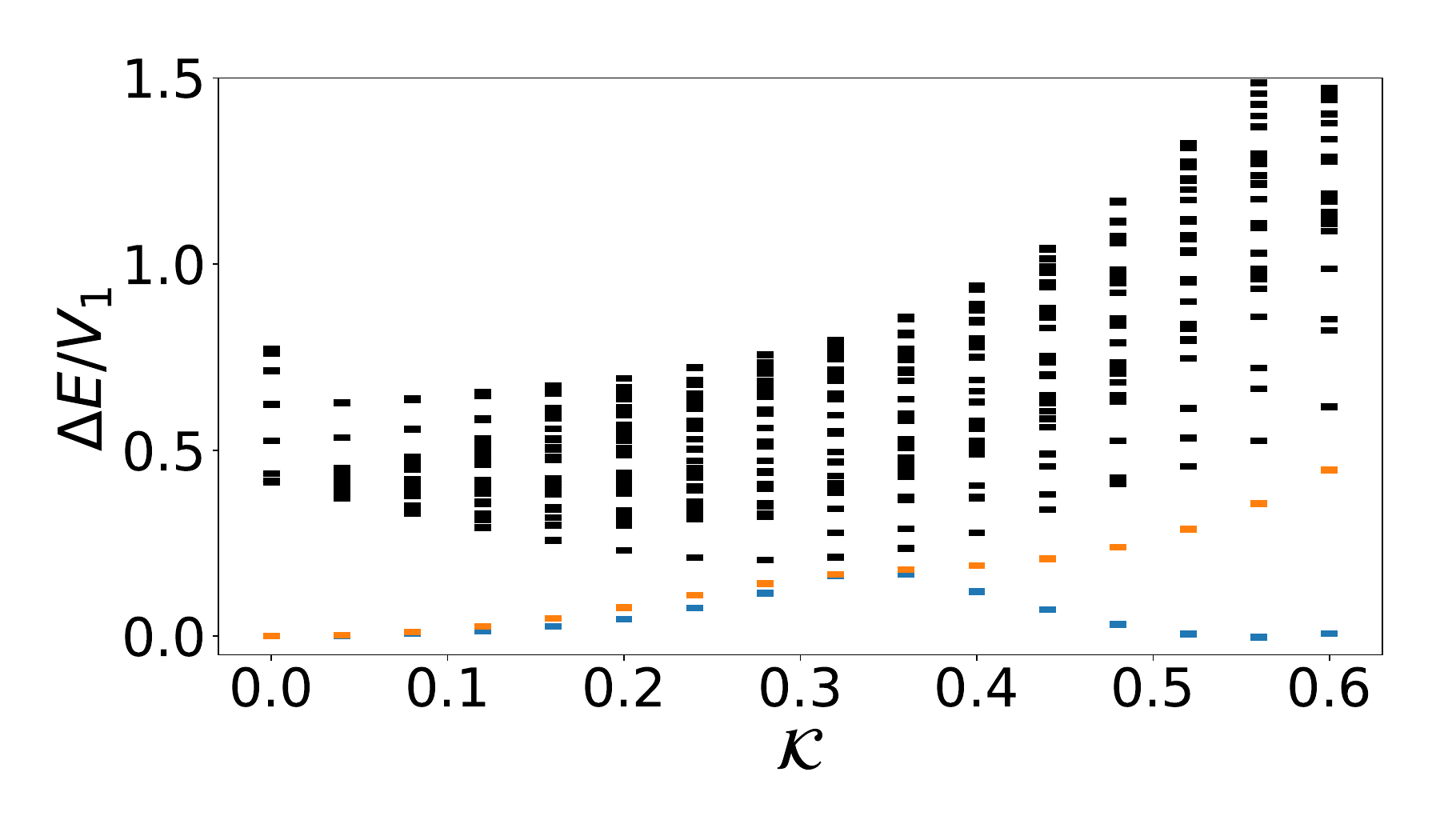}
    \caption{Evolution of the many-body spectrum at filling $2/3$ for $21$ unit cells.}
\label{fig:spectrum21sites}
\end{figure}

\begin{figure}
    \centering
    \includegraphics[width=\linewidth]{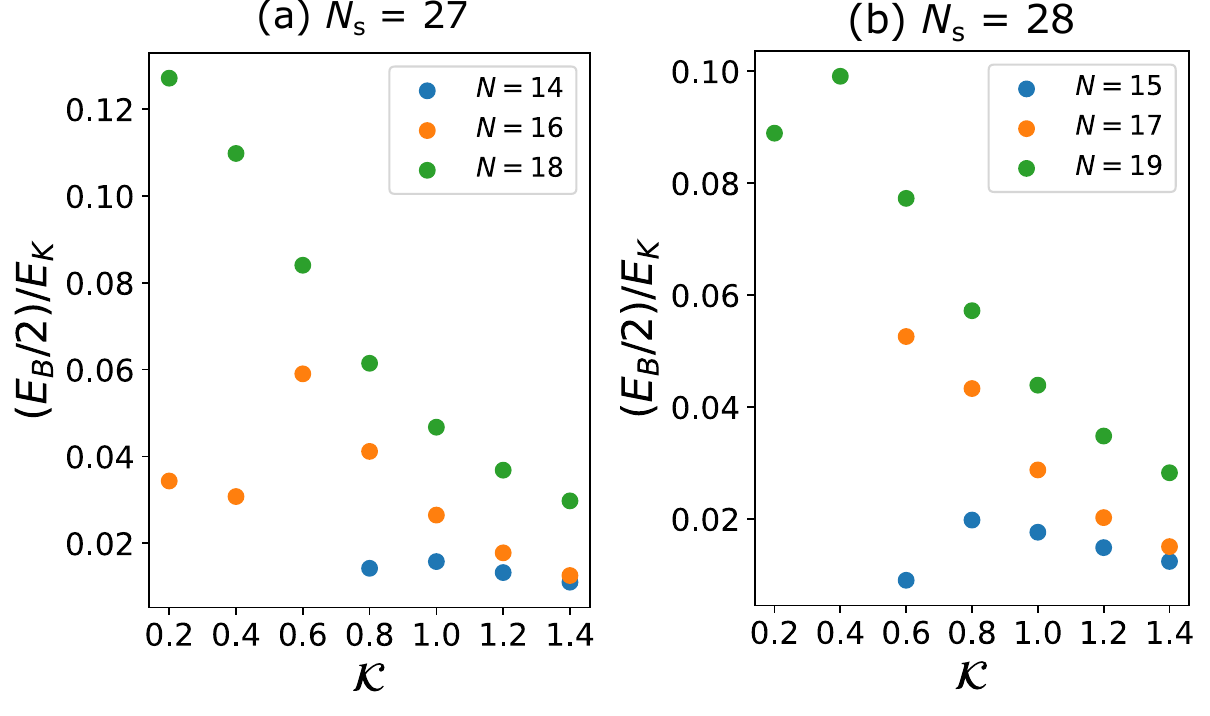}
    \caption{Ratio between binding energy and average kinetic energy per hole as a function of $\mathcal K$ doping for (a) $N_s = 27$ sites and (b) $N_s = 28$ sites for different number of particles.}
    \label{fig:binding_main}
\end{figure}

\section{Additional ED results}\label{app:ED}

In this section, we present additional numerical results supporting our theoretical findings, along with further details on the quantities shown in the main text.

We start by presenting Fig.~\ref{fig:spectrum21sites}, which depicts the evolution of the many-body ground state at filling factor $2/3$ for a $C_6$-symmetric cluster composed of $21$ unit cells. 
For this cluster geometry, the three FCIs are located at the three many-body momenta $\gamma$, $\kappa$ and $\kappa'$. 

\subsection{Binding energy}\label{app:EDBinding}

\begin{figure*}
    \centering
    \includegraphics[width=.65\linewidth]{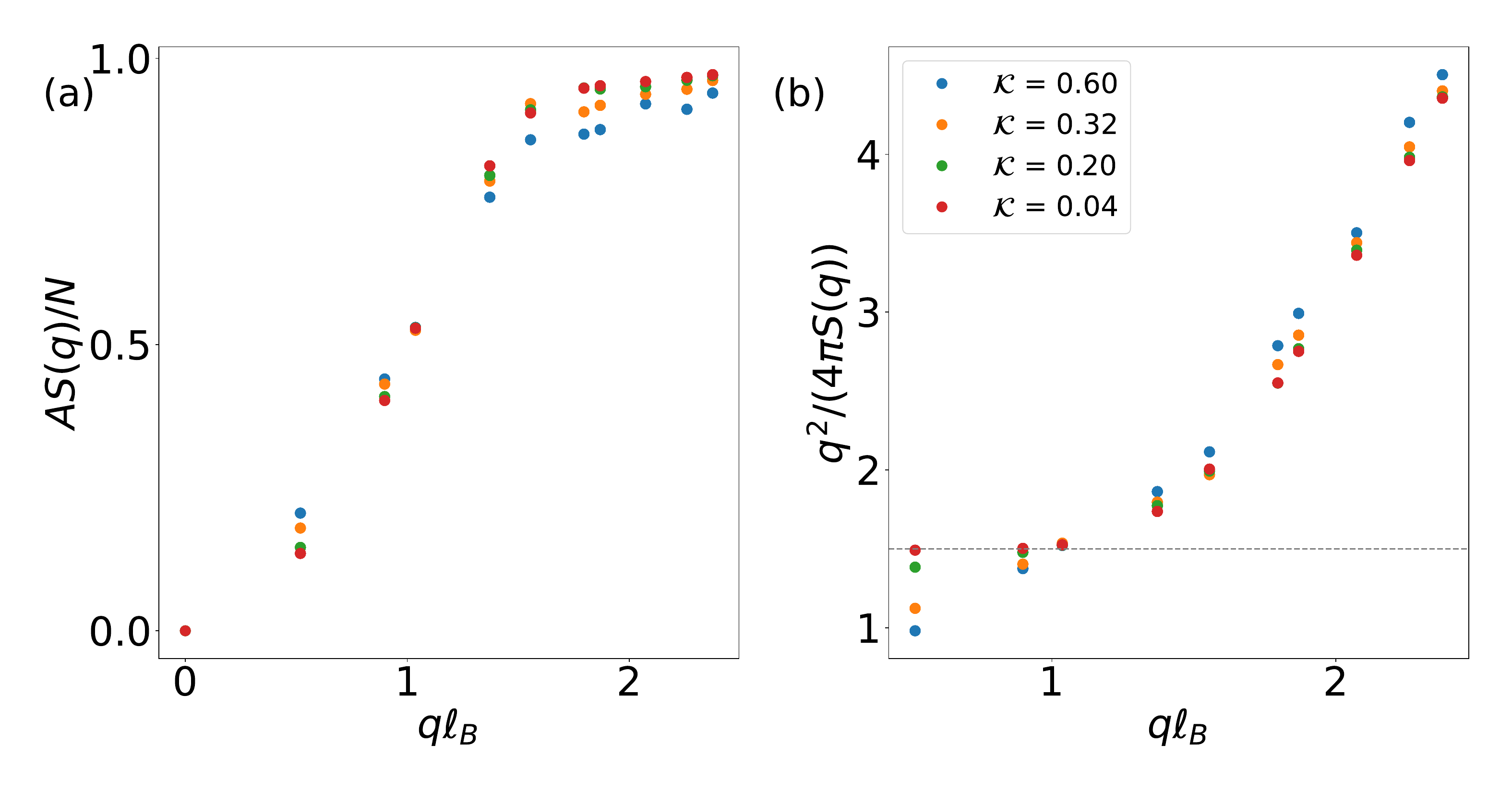}
    \caption{Evolution of the structure factor (Panel (a)) and $q^2/(4\pi S(\q))$ (Panel (b)) for different values of $\mathcal K$ at $\nu=2/3$ and $N_s=27$. }
    \label{fig:structure_factors_app}
\end{figure*}
We compute the pair binding energy
\begin{equation}\label{app:binding}
    E_B(N) = 2 E(N+1) - E(N+2) - E(N),
\end{equation}
where $N$ is the number of particles. 
Within our definition~\eqref{app:binding}, when $E_B>0$ this quantity measures the energetic gain associated with adding a charge-$2e$ excitation (a pair of particles) compared to adding two independent charge-$1e$ excitations.

We find that $E_B$ becomes positive, $E_B>0$, at filling $\nu=2/3$ and in its vicinity, indicating an effective attractive interaction between charge-$1e$ excitations. 
To characterize the strength of the superconducting pairing, we compute the ratio of the binding energy to the average kinetic energy of the holes in the many-body ground state  
\begin{equation}
    E_K  =\sum_{\k\in\rm BZ}\epsilon_{\k}\left( 1-\langle c^\dagger_{-\k} c_{-\k}\rangle\right)/N_h, 
\end{equation}
with $N_{h}$ the number of holes in the system and $\epsilon_{\k}$ the interaction induced dispersion defined below Eq.~\eqref{HartreeFock}. 

We observe that the ratio decreases with increasing $\mathcal{K}$ (Fig.~\ref{fig:binding_main}), indicating that the superconductor becomes more BCS-like, with the critical temperature $T_c$ set by the binding energy.
Conversely, decreasing $\mathcal{K}$, the ground state evolves toward a BEC-like regime, where $T_c$ is limited by the phase stiffness of the superconducting order. 


\subsection{Structure factor}

To characterize the transition between the two different regimes, we computed the full structure factor~\eqref{Sq}. 
To this aim, we first notice that
\begin{equation}
   {\langle\rho({\q})\rho({-\q})\rangle}=N+\sum_{\k_1\k_2} \langle f^\dagger_{\k_1}f^\dagger_{\k_2}f_{\k_2-\q}f_{\k_1+\q}\rangle,
\end{equation}
with $\k_1,\k_2$ unrestricted. Next, we express $\k$ as $\k=[\k]+\G$, where $[\k]$ belongs to the first Brillouin zone and $\G$ is a reciprocal lattice vector. 
Finally, in the flat Chern band basis the structure factor reads:
\begin{widetext}
\begin{equation}\begin{split}
    S(\q)=& \frac{N+\sum_{\k_1\k_2\in \rm BZ} \Lambda_{\k_1,[\k_1+\q]+\Q+\G_1}\Lambda_{\k_2,[\k_2-\q]-\Q+\G_2} \langle c^\dagger_{\k_1}c^\dagger_{\k_2}c_{[\k_2-\q]}c_{[\k_1+\q]}\rangle-\delta_{\q,\G}\left|\sum_{\k\in\rm BZ}\Lambda_{\k,\k+\G}n(\k)\right|^2}{A}.
\end{split}
\end{equation}
\end{widetext}
Fig.~\ref{fig:structure_factors_app}(a) shows the evolution of the structure factor for different values of $\mathcal K$ across the transition. 
On the other hand, Fig.~\ref{fig:structure_factors_app}(b) shows $q^2/(4\pi S(\q))$ which is related to the dispersion relation of the low-energy excitations obtained within the single mode approximation.

\section{Particle-hole transformation and interaction-induced dispersion}\label{app:PHdispersion}

The spatial modulation of the magnetic field has a profound impact~\eqref{hamiltonian}. 
In addition to reducing continuous magnetic translation symmetry to discrete translations, $\delta B(\r)$ explicitly breaks particle–hole symmetry, leading to distinct behaviors for electron and hole excitations. 
This is made explicity by performing the particle–hole transformation~\cite{abouelkomsan_particle-hole_2020}: $\mathcal P c_{\k}\mathcal P^\dagger =d_{-\k}^\dagger$.
Under this transformation, the electron filling factor $\nu$ maps to $1-\nu$, the Berry curvature $\Omega(\k)\to-\Omega(-\k)$, the magnetic field flips sign and the projected density operator for particles $\bar \rho(\q)$ transforms as: 
\begin{equation}\begin{split}\label{phdensityoperator}
    \mathcal P \bar\rho(\q) \mathcal P^\dagger
    =N_{s}-\bar n(\q),
\end{split}\end{equation}
with $\bar n(\q)=\sum_{\k\in\rm BZ}\braket{u_{-\k}}{u_{-\k-\q}}^*d^\dagger_{\k}d_{\k+\q}$ hole-like density operator and $N_{s}$ number of unit cells. 
In the hole quasiparticle basis, the Hamiltonian becomes: 
\begin{equation}\begin{split}
\label{phhamiltonian}
    \mathcal P H\mathcal P^\dagger  =& \sum_{\k\in\rm BZ} \epsilon_{\k} \left(d^\dagger_{\k}d_{\k}-1/2\right)\\
    &+\frac{1}{2A}\sum_{\q}V({\q}):\bar{n}({\q})\bar n({-\q}):,
\end{split}\end{equation}
where $\epsilon_{\k}=-\Sigma^{F}(-\k)-\Sigma^{H}(-\k)$ is the interaction-induced dispersion, i.e., the energy of a single hole doped into the Chern insulator at full filling of the Chern band ($\nu=1$). 
The dispersion includes two contributions, the Hartree and the Fock self-energies $\Sigma^H$ and $\Sigma^F$, respectively: 
\begin{equation}\begin{split}\label{HartreeFock}
    &\Sigma^F(\k)=-\frac{1}{A}\sum_{\k'\in\rm BZ}\sum_{\G}V({\k-\k'-\G})|\Lambda_{\k,\k'+\G}|^2,\\
    & \Sigma^H(\k)=\frac{1}{A}\sum_{\G}V(\G)\Lambda_{\k,\k+\G}\sum_{\k'\in \rm BZ}\Lambda_{\k',\k'-\G},
\end{split}\end{equation}
where the form factor $\Lambda_{\k,\k'+\G}$ is given in Eq.~\eqref{app:form_factors}.
 For $\mathcal K\neq 0$, the dispersion relation acquires a finite bandwidth $W$ increasing monotonously with $\mathcal K$~\cite{GuerciAbouelkomsan2025}.

\bibliography{biblio}

\end{document}